\documentclass[11pt]{article}
\pdfoutput=1
\usepackage{arxiv}

\usepackage{microtype}

\usepackage[table]{xcolor}
\usepackage[T1]{fontenc}
\usepackage{amsthm}
\usepackage{amssymb}
\usepackage[english]{babel}
\usepackage[utf8]{inputenc}
\usepackage{amsmath}
\usepackage{amsfonts}
\usepackage{graphicx}
\usepackage[colorinlistoftodos]{todonotes}
\usepackage{mathtools}
\usepackage{enumitem}
\usepackage{thmtools}
\usepackage{fullpage}
\usepackage{booktabs}
\usepackage{bbm}
\usepackage[normalem]{ulem} 
\usepackage[hidelinks]{hyperref}
\usepackage{multirow}

\newcommand{\N}{\mathbb{N}}
\newcommand{\R}{\mathbb{R}}
\newcommand{\Z}{\mathbb{Z}}
\newcommand{\Pc}{\mathcal{P}}

\newcommand{\A}{\mathbf{A}}

\newcommand{\E}{\mathbb{E}}

\newcommand{\vol}{\textrm{vol}}
\newcommand{\ABCD}{\textbf{ABCD}}
\newcommand{\hABCD}{\textbf{h--ABCD}}

\theoremstyle{plain}

\title{\textbf{H}ypergraph \textbf{A}rtificial \textbf{B}enchmark for \textbf{C}ommunity \textbf{D}etection (\hABCD) }

\author{
Bogumi\l{} Kami\'nski\thanks{Decision Analysis and Support Unit, SGH Warsaw School of Economics, Warsaw, Poland; e-mail: \texttt{bogumil.kaminski@sgh.waw.pl}}
\And
Pawe\l{}~Pra\l{}at\thanks{Department of Mathematics, Toronto Metropolitan University, Toronto, ON, Canada; e-mail: \texttt{pralat@ryerson.ca}. Part of this work was done while the author was visiting the Simons Institute for the Theory of Computing.}
\And
Fran\c{c}ois Th\'eberge\thanks{Tutte Institute for Mathematics and Computing, Ottawa, ON, Canada; email: \texttt{theberge@ieee.org}}
}

\begin{document}

\maketitle

\begin{abstract}
The \textbf{A}rtificial \textbf{B}enchmark for \textbf{C}ommunity \textbf{D}etection (\ABCD) graph is a recently introduced random graph model with community structure and power-law distribution for both degrees and community sizes. The model generates graphs with similar properties as the well-known \textbf{LFR} one, and its main parameter $\xi$ can be tuned to mimic its counterpart in the \textbf{LFR} model, the mixing parameter $\mu$. In this paper, we introduce \textbf{h}ypergraph counterpart of the \ABCD\ model, \hABCD, which also produces random hypergraph with distributions of ground-truth community sizes and degrees following power-law. As in the original \ABCD, the new model \hABCD\ can produce hypergraphs with various levels of noise. More importantly, the model is flexible and can mimic any desired level of homogeneity of hyperedges that fall into one community. As a result, it can be used as a suitable, synthetic playground for analyzing and tuning hypergraph community detection algorithms. 
\end{abstract}

\section{Introduction}

Many networks that are currently modelled as graphs would be more accurately modelled as hypergraphs. This includes the collaboration network in which nodes correspond to researchers and hyperedges correspond to papers that consist of nodes associated with researchers that co-authorship a given paper. After many years of intense research using graph theory in modelling and mining complex networks~\cite{newman2018networks,jackson2010social,easley2010networks,kaminski2021mining}, hypergraphs start gaining considerable traction~\cite{benson2016higher,benson2018simplicial,battiston2020networks,benson2021higher}. Standard but important questions in network science are revisited in the context of hypergraphs. This includes various aspects related to community detection in networks~\cite{kaminski2019clustering,kaminski2020community,Kumar1,Kumar2,chodrow2021generative,yin2017local,yin2018higher,benson2015tensor,chien2018community,ahn2018hypergraph} which we concentrate on in this paper. However, hypergraphs also create brand new questions which did not have their counterparts for graphs. For example, how hyperedges overlap in empirical hypergraphs~\cite{lee2021hyperedges}? Or how the existing patterns in a hypergraph affect the formation of new hyperedges~\cite{juul2022hypergraph}?

Despite the fact that currently there is a vivid discussion around hypergraphs, the theory and tools are still not sufficiently developed to allow most problems, including clustering, to be tackled directly within this context. Indeed, researchers and practitioners often create the 2-section graph of a hypergraph of interest (that is, replace each hyperedge with a clique) and apply classical tools designed for graphs. After moving to the 2-section graph, one clearly loses some information about hyperedges of size greater than two and so there is a common belief that one can do better by using the knowledge of the original hypergraph. 

As mentioned earlier, there are some recent attempts to directly deal with hypergraphs in the context of clustering. In~\cite{zhou2006learning}, methods from spectral clustering are generalized to hypergraphs; \cite{contisciani2022inference} proposes a framework to infer missing hyperedges and detect overlapping communities, while in~\cite{chodrow2022nonbacktracking}, extensions of non-backtracking spectral clustering are proposed in the context of hypergraphs. In Kumar et al.~\cite{Kumar1,Kumar2}, the authors reduce the problem to graphs but use original hypergraphs to iteratively adjust weights to encourage some hyperedges to be included in some cluster but discourage other ones (this process can be viewed as separating signal from noise). Moreover, in~\cite{kaminski2019clustering,kaminski2020community} a number of extensions of the classic null model for graphs are proposed that can potentially be used by hypergraph algorithms. 

Unfortunately, there are many ways such extensions can be done depending on how often nodes in one community share hyperedges with nodes from other communities. This is something that varies between networks at hand and usually depends on the hyperedge sizes. Let us come back to the collaboration network we discussed earlier. Hyperedges associated with papers written by mathematicians might be more homogeneous and smaller in comparison with those written by medical doctors who tend to work in large and  multidisciplinary teams. Moreover, in general, papers with a large number of co-authors tend to be less homogeneous, and other patterns can be identified~\cite{juul2022hypergraph}. A good clustering algorithm should be able to automatically decide which extension should be used. However, in order to be able to design and properly tune such algorithms, one needs to have a synthetic random hypergraph model that is able to simulate various scenarios.

The hypergraph model, \hABCD, is our response to this need from both practitioners and academia. This random graph model, similarly to the original \ABCD\ model, produces hypergraphs with community structure and power-law distribution for both degrees and community sizes. Indeed, as with graphs, power-law distribution is a common feature present in many real-world hypergraphs; for example, \cite{do2020structural} shows power-law distributions for several real-world hypergraph decompositions, and we have checked that in the presented cases the hypergraph degree distributions also follow power-law distributions.

\medskip

The paper is structured as follows. In Section~\ref{sec:existing_models}, we briefly discuss the history of \ABCD, the ``parent'' of the \hABCD\ model introduced in this paper, before summarizing other synthetic hypergraphs. We focus on models important in the context of community detection and explain the differences between the existing models and our model. The \hABCD\ model is defined in Section~\ref{sec:model}. Experiments highlighting important properties of the model are discussed in Section~\ref{sec:experiments}. In Section~\ref{sec:future}, we conclude the paper with a few future research directions. 

\section{Existing Models}\label{sec:existing_models}

\subsection{\ABCD\ graph Models}

There are very few datasets with ground-truth identified and labelled. As a result, there is need for synthetic random graph models with community structure that resemble real-world networks in order to benchmark and tune clustering algorithms that are unsupervised by nature. The \textbf{LFR} (\textbf{L}ancichinetti, \textbf{F}ortunato, \textbf{R}adicchi) model~\cite{lancichinetti2008benchmark,lancichinetti2009benchmarks} generates networks with communities and at the same time it allows for the heterogeneity in the distributions of both node degrees and of community sizes. It became a standard and extensively used method for generating artificial networks. 

The \textbf{A}rtificial \textbf{B}enchmark for \textbf{C}ommunity \textbf{D}etection (\ABCD)~\cite{kaminski2021artificial} was recently introduced and implemented\footnote{\url{https://github.com/bkamins/ABCDGraphGenerator.jl/}}, including a fast implementation\footnote{\url{https://github.com/tolcz/ABCDeGraphGenerator.jl/}} that uses multiple threads (\textbf{ABCDe})~\cite{kaminski2022abcde}. Undirected variant of \textbf{LFR} and \textbf{ABCD} produce graphs with comparable properties but \textbf{ABCD}/\textbf{ABCDe} is faster than \textbf{LFR} and can be easily tuned to allow the user to make a smooth transition between the two extremes: pure (disjoint) communities and random graph with no community structure. Moreover, it is easier to analyze theoretically---for example, in~\cite{kaminski2022modularity} various theoretical asymptotic properties of the \ABCD\ model are investigated including the modularity function that, despite some known issues such as the ``resolution limit'' reported in~\cite{fortunato2007resolution}, is an important graph property of networks in the context of community detection. Finally, the building blocks in the model are flexible and may be adjusted to satisfy different needs. For example, the original \ABCD\ model was recently adjusted to include potential outliers in~\cite{kaminski2022outliers} resulting in \textbf{ABCD+o} model. Adjusting the model to hypergraphs is much more complex but doable.

\subsection{Other Hypergraph Models}

The classical configuration model, which was first introduced by Bollob\'as~\cite{bollobas1980probabilistic}, is a standard model producing graphs with a given degree sequence. It was generalized to higher order structures many times~\cite{chodrow2020configuration,dyer2021sampling}. One early example is the folksonomy, a tripartite structure of users, resources, and tags, that was modelled as hypergraphs generated via configuration-type model in~\cite{ghoshal2009random}. The variant of the configuration model in~\cite{courtney2016generalized,courtney2017weighted} generalizes it even further, namely, to simplicial complexes. Simplicial complexes are attractive tools when studying topological aspects of discrete data~\cite{carlsson2009topology}, but subset inclusion is a strong property, not suitable from our application point of view, namely, community detection. In general, configuration models do not pay attention to labels of nodes and so cannot produce graphs with community structure. Having said that, they can be used as an ingredient of models which produce networks with communities. The configuration model was used in the original \ABCD\ model and we will use its generalization again for \hABCD.

\medskip

The Chung-Lu model for graphs (see~\cite{chung2006complex} for details) gives an efficient and simple way to generate graphs with an {\it expected} degree sequence, but without community structure. It can also be easily generalized for bipartite graphs, as described in~\cite{aksoy2017}; the bipartite representation is often used to model hypergraphs. 
In the same paper, a generalization of the {\bf BTER} (block two-level Erd\H{o}s-R\'enyi) model is proposed, also 
for bipartite graphs. In the standard BTER model, given degree distribution is preserved as well as the degree-wise clustering coefficients. With bipartite graphs, there are no 3-cycles, so a new {\it metamorphosis} coefficient is introduced, which is based on 4-cycles, and the model preserves this coefficient in a degree-wise matter.

\medskip

The stochastic block model (\textbf{SBM}), first introduced in~\cite{holland1983stochastic}, is one of the most important graph models in community detection and clustering. One benefit of it is that, being a generative model we can formally study the probability of inferring the ground truth. This is what distinguishes it from the configuration and Chung-Lu models. There are many variants and various applications of the \textbf{SBM}. For more details, we direct the reader to~\cite{abbe2017community}, one of the many surveys on this model. In particular, one can generalize the \textbf{SBM} to hypergraphs as it was done in, for example,~\cite{ghoshdastidar2017consistency,kim2018stochastic,brusa2022model}, while in~\cite{larremore2014}, a generalization of \textbf{SBM} is proposed for bipartite networks. Most SBM-type models have ground-truth communities embedded into the model but do not produce graphs with realistic degree distributions; a variant of SBM is proposed in~\cite{qiao2018adapting} to approximate power-law degree distribution but as far as we know, there is no such model for hypergraphs.

In~\cite{ruggeri2022principled}, an interesting model is introduced in which hyperedges can be generated via a sampling method, given the number of communities, node memberships, and affinity parameters between the communities. The node degrees and hyperedge sizes can either be given explicitly or sampled from the model. This approach is more complex than the one taken by the family of \textbf{ABCD} models. The authors of~\cite{ruggeri2022principled} report that generating sparse hypergraphs with $10^5$ nodes takes roughly half an hour, which is orders of magnitude slower than with \hABCD.

In~\cite{contisciani2022inference}, the probabilistic \textbf{Hypergraph-MT} model for hypergraphs is presented and is shown to be useful for inferring missing hyperedges as well as detecting overlapping communities. While the model as presented does not explicitly provide a benchmark, in the conclusion of the paper, the authors mention that their model can also be used to sample synthetic data with hypergraph structure.

Other hypergraphs models include core-periphery structure model~\cite{papachristou2022,tudisco2023core}, preferential attachment model with power-law degree distribution and high modularity~\cite{giroire2021}, and entropy-based models~\cite{saracco2022}.

Finally, the authors of~\cite{osti_1651312} propose the \textbf{HyGen} model that seems to be similar to the one introduced by us in this paper. Their algorithm is shown to be fast and scalable using MPI standard for its distributed generation. (MPI, message passing interface is a standardized means of exchanging messages between multiple computers running a parallel program across distributed memory.) The \textbf{HyGen} model uses the idea of building independently community and background hypergraphs that was also considered in the original \textbf{ABCD} graph generator~\cite{kaminski2022abcde} and is used in our \hABCD\ model. The main difference between the two models is that \textbf{HyGen} assumes that hyperedges belonging to communities are completely contained in them, that is, all members of a community hyperedge belong to one community. Our \hABCD\ algorithm allows for community hyperedges to be only partially contained within a community as long as majority of their members belong to that community. This feature is a significant advantage of \hABCD\ over \textbf{HyGen} as most complex networks have non-strict hyperedges. See Section~\ref{sec:model} for more details of community and background hypergraph generation.

\subsection{Distinctive Features of \hABCD}

The goal of this research project is to introduce (and efficiently implement) a scalable synthetic hypergraph benchmark model with community structure and power-law degree distribution as well as community sizes. Since none of the existing hypergraph models satisfy all of the desired properties, we moved back to graph models and tried to adjust one of them to our needs.

\hABCD\ model that is proposed in this paper produces networks that have power-law degree distributions and distributions of community sizes. Alternatively, the user may easily inject the two distributions as parameters of the model. The user may control the level of noise which covers all spectrum of possibilities of community strength. The model returns the ground-truth communities that then may be used to benchmark and tune community detection algorithms. The model, by design, is fast but it is also efficiently implemented in Julia language. (Julia is a high-level, high-performance, dynamic programming language that recently gains a lot of interest in scientific computing applications~\cite{bezanson2017julia}.) As a result, it is by orders of magnitude faster than other models that we are aware of. Finally, because of its simplicity, its properties as well as various processes (such as spreading of information, anomalies detection, etc.) can be analyzed theoretically. Such studies, complemented by simulations that are typically easier to perform, often uncover important mechanisms that are present in real-world complex networks. One such spectacular example is a study of the preferential attachment model that explains the following phenomena: ``rich-get-richer'' mechanisms are responsible for creating power-law distributions that are typically observed in complex networks~\cite{barabasi1999emergence}. Analysis of theoretical properties of \textbf{h-ABCD} are left as future work.

\section{Definition of the Model} \label{sec:model}

\subsection{Parameters of the Model}

Let us start with introducing parameters of the model. The \hABCD\ model is governed by parameters summarized in Table~\ref{tab:parameters}. Note, in particular, that the number of hyperedges, $m$, is not a parameter of the model. It is a random variable that depends on the number of nodes $n$, the degree distribution of the graph, and the distribution of hyperedge sizes $q_d$. This random variable is well concentrated around its expectation but it is not a parameter provided as an input. Similarly, the number of communities, $\ell$, is not a parameter of the model but a random variable that depends on the number of nodes $n$ and the distribution of community sizes. The process of determining of these values, as well as a detailed explanation of the interpretation of the presented parameters, is discussed further in this section.

\begin{table}[htp]
\begin{center}
\begin{tabular}{|l|l|p{0.6\linewidth}|}
\hline
parameter & range & description \\
\hline
$n$ & $\N$ & number of nodes \\
$\gamma$ & $(2,3)$ & power-law exponent of degree distribution \\
$\delta$ & $\N$ & minimum degree at least $\delta$ \\
$D$ & $\N$ & maximum degree at most $D$ \\
$\beta$ & $(1,2)$ & power-law exponent of distribution of community sizes \\
$s$ & $\N \setminus [\delta]$ & community sizes at least $s$ \\
$S$ & $\N \setminus [s-1]$ & community sizes at most $S$ \\
$\xi$ & $(0,1)$ & level of noise (fraction of non-community hyperedges) \\
$L$ & $\N$ & size of largest hyperedges  \\
$q_d$ & $[0,1]$ & fraction of hyperedges that are of size $d$; $\sum_{d=1}^Lq_d=1$ \\
$w_{c,d}$ & $[0,1]$ & fraction of community hyperedges of size $d$ that have exactly $c$ within-community nodes; $\sum_{c=\lfloor d/2 \rfloor + 1}^dw_{c,d}=1$ \\
\hline
\end{tabular}
\end{center}
\caption{Parameters of the \hABCD\ model}
\label{tab:parameters}
\end{table}

The model generates a hypergraph on $n$ nodes. The degree distribution follows power-law with exponent $\gamma$, minimum and maximum value equal to $\delta$ and, respectively, $D$. Community sizes are between $s$ and $S$, and also follow power-law distribution, but this time with exponent $\beta$. The suggested range of values for parameters $\gamma$ and $\beta$ are chosen according to experimental values commonly observed in complex networks not only represented as graphs~\cite{barabasi2016network,orman2009comparison} but also as hypergraphs~\cite{do2020structural}. Parameter $\xi$ is responsible for the level of noise. If $\xi = 0$, then each hyperedge is a community hyperedge meaning that majority of its nodes belong to one community. On the other extreme, if $\xi=1$, then communities do not play any roles and hyperedges are simply ``sprinkled'' across the entire hypergraph that we will refer to as background hypergraph. Vector $(q_1, \ldots, q_L)$ determines the distribution of the number of hyperedges of a given size. 

Finally, parameters $w_{c,d}$ specify how many nodes from its own community a given community hyperedge should have. We call a community hyperedge to be of a $(c,d)$ type if it has size $d$ and exactly $c$ of its nodes belong to one of the communities. Note, in particular, that we require that a community hyperedge must have more than a half of its nodes from the community. Therefore, $w_{c,d}$ is defined for $d/2<c\leq d$, where $d\in[L]$.

The model is flexible and may accept any family of parameters $w_{c,d}$ satisfying specific needs of the users, but here is a list of three standard options implemented in the code (see Table~\ref{tab:wcd} for example calculations): 
\begin{itemize}
    \item {\em majority} model: $w_{c,d}$ is uniform for all admissible values of $c$, that is, for any $d/2<c\leq d$,
    $$w_{c,d} = \frac {1}{(d - \lfloor d/2 \rfloor)} = \frac {1}{\lceil d/2 \rceil},$$
    \item {\em linear} model: $w_{c,d}$ is proportional to $c$ for all admissible values of $c$, that is, for any $d/2<c\leq d$,
    $$w_{c,d} = \frac {2c}{(d+\lfloor d/2 \rfloor + 1)(d - \lfloor d/2 \rfloor)} = \frac {2c}{(d+\lfloor d/2 \rfloor + 1) \lceil d/2 \rceil},$$
    \item {\em strict} model: only ``pure'' hyperedges are allowed, that is
    $$w_{d,d}=1 \quad\text{and}\quad w_{c,d}=0 \quad\text{for}\quad d/2 < c < d.$$
\end{itemize}

\noindent
In the above formulas and later in the paper, for a given $x \in \R$, $\lfloor x\rfloor$ is used to denote the floor of $x$ (that is, the largest integer not larger than $x$) and $\lceil x\rceil$ to denote its ceiling (that is, the smallest integer not smaller than $x$).

\definecolor{Gray}{gray}{0.9}
\newcolumntype{?}{!{\vrule width 1pt}}

\makeatletter
\def\hlinewd#1{%
\noalign{\ifnum0=`}\fi\hrule \@height #1 \futurelet
\reserved@a\@xhline}
\makeatother

\begin{table}
\begin{tabular}{|c?c|c|c|c|c|}
\multicolumn{6}{c}{\emph{majority} $w_{c,d}$} \\
\hline
\multirow{2}{*}{$c$} & \multicolumn{5}{c|}{$d$} \\ \cline{2-6} 
  & 1 & 2 & 3 & 4 & 5\\ \hlinewd{1pt}
1 & 1 & \cellcolor{Gray} & \cellcolor{Gray} & \cellcolor{Gray} & \cellcolor{Gray} \\ \hline
2 & \cellcolor{Gray} & 1 & 1/2  & \cellcolor{Gray} & \cellcolor{Gray} \\ \hline
3 & \cellcolor{Gray} & \cellcolor{Gray} & 1/2  & 1/2 & 1/3 \\ \hline
4 & \cellcolor{Gray} & \cellcolor{Gray} & \cellcolor{Gray} & 1/2 & 1/3 \\ \hline
5 & \cellcolor{Gray} & \cellcolor{Gray} & \cellcolor{Gray} & \cellcolor{Gray} & 1/3 \\ \hline
\end{tabular}
\hfill
\begin{tabular}{|c?c|c|c|c|c|}
\multicolumn{6}{c}{\emph{linear} $w_{c,d}$} \\
\hline
\multirow{2}{*}{$c$} & \multicolumn{5}{c|}{$d$} \\ \cline{2-6} 
  & 1 & 2 & 3 & 4 & 5\\ \hlinewd{1pt}
1 & 1 & \cellcolor{Gray} & \cellcolor{Gray} & \cellcolor{Gray} & \cellcolor{Gray} \\ \hline
2 & \cellcolor{Gray} & 1 & 2/5  & \cellcolor{Gray} & \cellcolor{Gray} \\ \hline
3 & \cellcolor{Gray} & \cellcolor{Gray} & 3/5  & 3/7 & 3/12 \\ \hline
4 & \cellcolor{Gray} & \cellcolor{Gray} & \cellcolor{Gray} & 4/7 & 4/12 \\ \hline
5 & \cellcolor{Gray} & \cellcolor{Gray} & \cellcolor{Gray} & \cellcolor{Gray} & 5/12 \\ \hline
\end{tabular}
\hfill
\begin{tabular}{|c?c|c|c|c|c|}
\multicolumn{6}{c}{\emph{strict} $w_{c,d}$} \\
\hline
\multirow{2}{*}{$c$} & \multicolumn{5}{c|}{$d$} \\ \cline{2-6} 
  & 1 & 2 & 3 & 4 & 5\\ \hlinewd{1pt}
1 & 1 & \cellcolor{Gray} & \cellcolor{Gray} & \cellcolor{Gray} & \cellcolor{Gray} \\ \hline
2 & \cellcolor{Gray} & 1 & 0  & \cellcolor{Gray} & \cellcolor{Gray} \\ \hline
3 & \cellcolor{Gray} & \cellcolor{Gray} & 1  & 0 & 0 \\ \hline
4 & \cellcolor{Gray} & \cellcolor{Gray} & \cellcolor{Gray} & 1 & 0 \\ \hline
5 & \cellcolor{Gray} & \cellcolor{Gray} & \cellcolor{Gray} & \cellcolor{Gray} & 1 \\ \hline
\end{tabular}
\caption{Example values of $w_{c,d}$ for \emph{majority}, \emph{linear}, and \emph{strict} weights for $d\in[5]$.}\label{tab:wcd}
\end{table}

\subsection{The Big Picture}

We summarize the main steps followed to generate the hypergraph \hABCD\ below. More details are provided right after.

\begin{enumerate}
    \item Sample degrees of nodes.
    \item Sample community sizes ensuring that the desired properties hold, in particular, that the sum of community sizes equals the number of nodes $n$.
    \item Compute the number of hyperedges of size one and then generate them.
    \item For each node, compute what fraction of its degree should be assigned to community hyperedges and what fraction remains to be used for background hyperedges.
    \item Assign nodes to communities ensuring that they fit into them (for example, nodes of very high degree cannot be assigned to small communities, as they would not be able to form simple hyperedges within such communities). One of such admissible assignments is selected randomly.
    \item Generate community hypergraphs.
    \item Generate the background hypergraph.
    \item Resolve potential problems with infeasible hyperedges (either being multisets or duplicates) by executing the rewiring process.
\end{enumerate}

\subsection{Definition of the Model---Details}

\subsubsection*{Simple Hypergraphs vs.\ Multi-hypergraphs}

Two variants of the model are considered in this paper, and both of them are implemented and available at the associated GitHub repository\footnote{\url{https://github.com/bkamins/ABCDHypergraphGenerator.jl}}. The first variant (that is assumed to be used by default) produces simple hypergraphs whereas the second one generates multi-hypergraphs. ``Simple'' in the context of hypergraphs means that hyperedges are sets of nodes but multi-sets are not allowed. Indeed, since all edges in a graph are of size two, loops in the context of graphs are multi-sets of size two in which one node is repeated twice. In particular, there are no multi-sets of size two which correspond to loops in the graph terminology. Similarly, no hyperedges can be repeated so, in particular, there are no parallel edges in the language of graphs (two identical hyperedges of size two in the language of hypergraphs).

\subsubsection*{Degree Distribution}

In the context of hypergraphs, the degree of a node $v \in V$ of a hypergraph $G=(V,E)$ is defined to be the number of hyperedges this node belongs to, regardless of their sizes. (For multi-hypergraphs, the number of occurrences of a node in a hyperedge as well as the number of repetitions of a hyperedge are taken into account.) The volume $\vol(V)$ of $G$ is defined to be the sum of degrees over all nodes in the hypergraph. Hence, if there are $m_d$ hyperedges of size $d$, then $\vol(V) = \sum_{d=1}^L d \, m_d$ giving us the counterpart of the \emph{handshaking lemma} for hypergraphs. Also denote $m = \sum_{d=1}^L m_d$ to be a total number of hyperedges in the hypergraph. As noted above, in \hABCD, in contrast to $n$ (the number of nodes), $m$ is a random variable, not a parameter of the model.

The degree distribution of nodes of \hABCD\ is generated randomly following the (truncated) \emph{power-law distribution} $\Pc(\gamma, \delta, D)$ with exponent $\gamma \in \R_+$, minimum value $\delta \in \N$, and maximum value $D \in \N$ ($\delta \le D$). Formally, if $X \in \Pc(\gamma, \delta, D)$, then for any $k \in \{ \delta, \delta+1, \ldots, D\}$ we have that 
\begin{equation}\label{eq:power-law-discrete}
\Pr( X = k ) = \frac { k^{-\gamma} }{ \sum_{x = \delta}^{D} x^{-\gamma} }.
\end{equation}
For typical applications, it is recommended to use $\gamma \in (2,3)$~\cite{barabasi2016network,orman2009comparison}, some small value of $\delta$ such as 5 or 10, and $D \approx n^{\zeta}$ for some $\zeta \in (0,1)$, where $n$ is the number of nodes. 

In order to generate \hABCD\ hypergraph on $n$ nodes with a given degree distribution $\textbf{x} := (x_1, \ldots, x_n)$, we will use a straightforward generalization of the classical random graph model with a given degree sequence known as the \textbf{configuration model} (sometimes called the \textbf{pairing model}), which was first introduced by Bollob\'as~\cite{bollobas1980probabilistic}. (See~\cite{bender1978asymptotic,wormald1984generating,wormald1999models} for related models and results.) We start with a set $P$ of $\vol(V) = \sum_{i=1}^n x_i$ \emph{points} that is partitioned into $n$ \emph{buckets} labelled with labels $v_1, \ldots, v_n$; bucket $v_i$ consists of $x_i$ points. We will additionally randomly partition the set $P$ into disjoint sets $P_h$, $h\in[m]$, of various sizes such that $P = P_1 \cup \ldots \cup P_m$, and construct a multi-hypergraph $\Pc(\textbf{x})$ as follows: nodes are the buckets $v_1, \ldots, v_n$, and a set $P_h$ of points corresponds to a hyperedge $e_h \subseteq V$ (possibly a multi-set) in $\Pc(\textbf{x})$ that consists of nodes (buckets) that contain some point in $P_h$. In our model, the process of generating random partition $P = P_1 \cup \ldots \cup P_m$ is quite complex and will be done in various phases. At this stage let us only mention that eventually there will be $m_d$ sets/hyperedges of size $d$ for a total of $m$ hyperedges. We will provide more details soon.

\subsubsection*{Distribution of Community Sizes}

Community sizes of \hABCD\ are generated randomly following the (truncated) \emph{power-law distribution} $\Pc(\beta, s, S)$ with exponent $\beta \in \R_+$, minimum value $s \in \N \setminus [\delta]$, and maximum value $S \in \N$ ($s \le S$). It is recommended to use $\beta \in (1,2)$~\cite{barabasi2016network,orman2009comparison}, some relatively small value of $s$ (in our experiments in this paper we set $s=50$), and $S \approx n^{\tau}$ for some $\zeta < \tau < 1$. The assumption that $\tau > \zeta$ is recommended to make sure large degree nodes have large enough communities to be assigned to. Similarly, the assumption that $s \ge \delta+1$ is required to guarantee that small communities are not too small and so that they can accommodate small degree nodes. (These conditions are needed to make sure that generating a simple hypergraph with the desired properties is feasible.)

The procedure of sampling the community size distribution is as follows. First, it is checked if for a given parameters $s$, $S$, and $n$ it is possible to generate appropriate community sizes. If it is not possible, then the generation process is stopped and error is returned. Then, the algorithm draws \texttt{maxiter} samples from the truncated power-law distribution; in the implementation, \texttt{maxiter}=1{,}000. Each sample is a vector $(c_1, \ldots, c_\ell)$ of community sizes such that $\sum_{j \in [\ell]} c_j \ge n$ and $\sum_{j \in [\ell-1]} c_j < n$. If for any of these samples the corresponding sum is exactly $n$, then the process is stopped and the obtained community size distribution is retained. If none of the \texttt{maxiter} samples yields the sum equal to $n$, then from the obtained distributions a vector $(c_1, \ldots, c_\ell)$ with minimum sum is selected; note that $\sum_{j \in [\ell]} c_j > n$. Clearly, $\ell > n/S$ but, more importantly, $\ell \le \lceil n/s \rceil$. The algorithm checks if $\ell > n/s$, and in such (highly unlikely) case we truncate the vector to that length; note that then the sizes add up to less than $n$. The reason for checking this condition is that it is impossible to generate a distribution having the sum equal to $n$, length greater than $n/s$, and with all entries at least $s$. Finally, we start the process of fixing the community sizes until their sizes add up to $n$. The updates are made in rounds. In each round all community sizes are shuffled in random order. Then, sequentially, their sizes are increased or decreased by $1$, depending if the total sum of sizes is less than or greater than $n$. To satisfy the desired properties, for a given community an update is made only if it results in a new community size in the range from $s$ to $S$. The process stops if the sum of community sizes reaches $n$. If the sum does not reach $n$ in one round, then we repeat the process until the desired property is reached.

\subsubsection*{Distribution of Hyperedge Sizes and Their Compositions}

The distribution of hyperedge sizes is captured by a vector $(q_1, \ldots, q_L)$ with $\sum_{d=1}^{L} q_d = 1$. The value of $q_d \in [0,1]$ indicates what fraction of the total volume is devoted to form hyperedges of size $d$. The model distinguishes two types of hyperedges: community hyperedges and background ones. Community hyperedges, by design, will have majority of their members to be part of one community. Because of the majority rule, each community hyperedge is uniquely assigned to one such community. The distribution of the desired composition of hyperedges is reflected by parameters $w_{c,d}$ with $d \in [L]$ and $d/2 < c \le d$ such that for each $d \in [L]$, $\sum_{c=\lfloor d/2 \rfloor+1}^d w_{c,d}= 1$. The value of $w_{c,d}\in[0,1]$ guides what fraction of community hyperedges of size $d$ have exactly $c$ members from its own community. We will call such hyperedges to be of type $(c,d)$. 

\subsubsection*{Hyperedges of Size 1}

Hyperedges of size 1 play a special role and therefore are generated before hyperedges of larger sizes. For example, they are ``neutral'' for community detection algorithms, such as the classical Louvain algorithm that uses the modularity function, since they are inherently part of communities their unique node belongs to. Having said that, potential users of the \hABCD\ model might be interested in generating such hyperedges for some other reasons, and so we provide such option.

The number of hyperedges of size 1 is $m_1 = \lfloor q_1 \vol(V) \rceil$, where for a given integer $a \in \Z$ and real number $b \in [0, 1)$ the random variable $\lfloor a+b \rceil$ is defined as 
\begin{equation}\label{eq:rounding}
\lfloor a+b \rceil = 
\begin{cases}
a & \text{ with probability } 1-b \\
a+1 & \text{ with probability } b.
\end{cases}
\end{equation}
(Note that $\E [\lfloor a+b \rceil] = a(1-b) + (a+1)b = a + b$.) In the variant of the model generating multi-hypergraphs, each hyperedge selects a node with probability proportional to the number of points in the configuration model associated with this node that are not assigned to any hyperedges yet. In order to generate simple hypergraphs, only points associated with nodes that are not yet assigned to any hyperedges are considered. For this to be feasible, a trivial condition has to be satisfied: $m_1 \le n$. If this property does not hold, then we truncate it to $m_1=n$ and, as a result, each node has exactly one hyperedge of size 1 that it is a part of. Finally, regardless whether we generate multi-hypergraphs or simple ones, we update the degree distribution $(x_1, \ldots, x_n)$ to reflect the fact that some nodes are associated with hyperedges of size 1. In other words, in the following subsections it will be convenient to assume that the degree distribution $(x_1, \ldots, x_n)$ ignores hyperedges of size one but only with them the degree distribution matches the one requested by the user.

\subsubsection*{Level of Noise} 

As mentioned earlier, there are two types of hyperedges of size at least 2: community hyperedges and background hyperedges. The ratio between the two types is guided by the main parameter of the model: $\xi \in [0,1]$. Indeed, the expected fraction of points (ignoring the ones that are already associated with hyperedges of size 1) that are going to be associated with background hyperedges is equal to $\xi$. In order to achieve this desired property, we split the degree $x_i$ of each node (equivalently, the associated points in the configuration model) into community degree $y_i$ and background degree $z_i$ ($x_i = y_i + z_i$). We split the weights randomly as follows: for each node, we independently assign $z_i = \lfloor \xi x_i \rceil$ and fix $y_i = x_i - z_i$ (or, equivalently, assign $y_i = \lfloor (1-\xi) x_i \rceil$ and fix $z_i = x_i - y_i$).

Recall that the degree of $v$ is equal to the number of hyperedges $v$ belongs to (in case of non-simple hypergraphs, taking into account both potential repetitions of hyperedges as well as repetitions of nodes within one hyperedge). On the other hand, neighbours of node $v$ are defined to be the nodes that are together with $v$ in some hyperedge. (Alternatively, one may consider the so-called 2-section multi-graph in which each hyperedge of size $d$ is replaced by a complete graph on $d$ nodes from that hyperedge. Then the neighbours of $v$ in the hypergraph are simply neighbours in the associated 2-section graph.) Since hyperedge sizes could be larger than 2, the number of neighbours of $v$ is often larger than the degree of $v$. In the original \ABCD\ model for graphs, parameter $\xi$ guided the fraction of the degree of a node $v$ that is assigned to non-community edges which coincided with the fraction of neighbours of $v$ from community that is different than the community of $v$. For hypergraphs, in the \hABCD\ model introduced in this paper, parameter $\xi$ still guides the fraction of the degree of $v$ that is assigned to background hyperedges but it does \emph{not} have a direct interpretation for the number of neighbours of $v$ that belong to different communities than $v$. For hypergraphs, this fraction will be generally larger than $\xi$. Indeed, even in the extreme case when $\xi=0$ (that is, all hyperedges are community hyperedges), hyperedges of size greater than $2$ that are assigned to some community will have majority of their members from such community but may not have all of them, depending on the hyperedge type distribution. Unless there are only hyperedges of type $(d,d)$, we still should expect some nodes forming these hyperedges to be from outside of such communities. As a result, not all neighbours of node $v$ will belong to the community of $v$.

\subsubsection*{Assigning Nodes into Communities}

At this point, the degree distribution ($x_1 \ge x_2 \ge \ldots \ge x_n$) and the distribution of community sizes ($c_1 \ge c_2 \ge \ldots \ge c_{\ell}$) are already fixed. (In fact, the degree distribution is already split into the community degree distribution ($y_i$, $i\in [n]$) and the background degree distribution ($z_i$, $i \in [n]$).) In what follows, as already signalled above, we assume that both the degree sequence $(x_i)$ as well as the sequence of community sizes $(c_j)$ are sorted in a non-increasing order.

\hABCD\ hypergraph will be formed as the union of $\ell+1$ independent hypergraphs: $\ell$ community hypergraphs $G_j=(V,E_j)$ in which $C_j \subseteq V$ with $|C_j|=c_j$ plays a special role ($j \in [\ell]$), and a single background hypergraph $G_0=(V,E_0)$, where $V = \bigcup_{j \in [\ell]} C_j$. Recall that a hyperedge belongs to a community if majority of its nodes belong to it. By design, all hyperedges of $G_j$, $j \in [\ell]$, (community hyperedges) will belong to its own community (which does not mean that all members of these hyperedges belong to $C_j$; that is why $V(G_j)=V$ instead of $V(G_j)=C_j$) but a few additional edges from the background graph $G_0$ might end up in that community. In order to create enough room for these hyperedges, node of degree $x_i$ needs to be assigned to community $C_j$ of large enough size $c_j$. The property that needs to be satisfied is as follows: for any $i \in [n]$, node $v_i$ is allowed to be assigned to a community of size $c_j$ if for any $d \in [L] \setminus \{1\}$ and any $\lfloor d/2 \rfloor + 1 \le c \le d$,
\begin{align}
& y_i\left(\sum_{f=\lfloor d/2 \rfloor +1}^c \, q_d \, w_{f,d} \cdot \binom{d-f}{c-f} \left( \frac {c_j}{n} \right)^{c-f} \left( 1 - \frac {c_j}{n} \right)^{d-c}\right) \nonumber \\
& + z_i \, q_d \cdot \binom{d-1}{c-1} \left( \frac {c_j}{n} \right)^{c-1} \left( 1 - \frac {c_j}{n} \right)^{d-c}   \le \binom{ c_j-1 }{ c-1 } \binom{n-c_j}{d-c}. \label{eq:prop_community_assignment}
\end{align}
This condition is especially important for the variant which generates simple hypergraphs but we insist on it even when multi-hypergraphs are created to avoid creating hyperedges in which nodes are repeated multiple times. For example, if a hyperedge of size 100 that is of type $(100,100)$ is assigned to a community of size 10, then some node will have to be repeated at least 10 times.

The rationale behind the above inequality is as follows. There are $y_i$ points associated with community degree of $v_i$ and we expect $q_d w_{f,d}$ fraction of them to be devoted to community hyperedges of type $(f,d)$. Each of these hyperedges has a probability of having precisely $c$ members in community $C_j$ well approximated by
$$
\binom{d-f}{c-f} \left( \frac {\vol(C_j)}{\vol(V)} \right)^{c-f} \left( 1 - \frac {\vol(C_j)}{\vol(V)} \right)^{d-c}.
$$ 
(Unfortunately, to verify this, the reader needs to wait for the description of the processes that generate community and background graphs.) The volume of $C_j$ is not known before the assignment of nodes into communities is finalized but, assuming that nodes are assigned randomly, it is expected that $\vol(C_j) = (c_j / n) \vol(V)$. 
Similarly, $z_i q_d$ points associated with background degree of $v_i$ is expected to be devoted to background hyperedges of size $d$. Each of these points has a probability of having precisely $c$ members in community $C_j$ well approximated by
$$
\binom{d-1}{c-1} \left( \frac {\vol(C_j)}{\vol(V)} \right)^{c-1} \left( 1 - \frac {\vol(C_j)}{\vol(V)} \right)^{d-c}.
$$ 
Hence, the left hand side of~(\ref{eq:prop_community_assignment}) corresponds to the expected number of hyperedges of type $(c,d)$ from both community $C_j$ and the background graphs. This explains~(\ref{eq:prop_community_assignment}) since in simple hypergraphs the community of size $c_j$ may accommodate at most $\binom{c_j-1}{c-1} \binom{n-c_j}{d-c}$ hyperedges of type $(c,d)$ containing node $v_i$. 

Let us also briefly comment on the complexity aspect of this part of the algorithm, since it is one of the two computationally expensive parts of the algorithm. Note that~(\ref{eq:prop_community_assignment}) can be rewritten in the form $y_i A_j + z_i B_j \le C_j$, where constants $A_j, B_j, C_j$ depend on the distribution of community sizes but does not depend on the degree distribution. As a result, these constants can be pre-computed for all triples $(j,d,c)$, where $j \in [\ell]$, $d \in [L] \setminus \{1\}$, and $\lfloor d/2 \rfloor + 1 \le c \le d$, before verifying the inequality. Moreover, for any group of nodes of degree $x_i$ with the same split into $y_i$ and $z_i$ we need to check the condition (\ref{eq:prop_community_assignment}) only once. These observations significantly reduce the running time of the algorithm.

An assignment of nodes into communities will be called admissible if the above family of inequalities~(\ref{eq:prop_community_assignment}) is satisfied for all nodes. Our goal is to select one admissible assignment at random, with distribution close to being uniform. Sampling uniformly one of such assignments for graphs turns out to be relatively easy from both theoretical and practical points of view~\cite{kaminski2021artificial,kaminski2022modularity}. Indeed, in order to assign nodes to communities, the following easy and natural algorithm is used. We consider nodes, one by one, in non-increasing order of their degrees, that is, we start with $v_1$ (highest degree node) and finish with $v_n$ (lowest degree node). We assign node $v_i$ randomly to one of the communities that have size large enough so that~(\ref{eq:prop_community_assignment}) is satisfied, and still have some ``available spots.'' We do it with probability proportional to the number of available spots left.

It is easy to see that for graphs the above simple algorithm generates one of the admissible assignments uniformly at random. The proof uses the fact that if node $v_i$ can be assigned to a given community, then node $v_{i'}$ for some $i'>i$ can also be assigned to that community. For hypergraphs, each node has to satisfy a few inequalities~(\ref{eq:prop_community_assignment}) (for various combinations of $c$ and $d$) and so this property might not be satisfied. However, this simple algorithm still produces near uniform sampling that is statistically indistinguishable from uniform, provided that $n$ is large enough.

Note that a randomly selected community (with probability proportional to the number of available spots left) often satisfies inequalities~(\ref{eq:prop_community_assignment}), especially later in the process when nodes of small degrees are being assigned. Hence, in order to speed up the generation process, we first select 10 random communities and only if none of them satisfies the desired property, we move on to the exhaustive search that identifies potential destinations before selecting one of them randomly.

\subsubsection*{Creating Community Graphs}

In order to generate the community hypergraphs $G_j = (V, E_j)$ we will use a generalization of the configuration model. Recall that there are $p_j = \sum_{v_i \in C_j} y_i$ points associated with nodes in $C_j$ that will be part of community hyperedges. Some of such points will belong to community hyperedges of type $(c,d)$ with $c > d/2$ members from $C_j$ but some of them will form hyperedges with majority of their members from some other community.

First, we need to fix the number of hyperedges of a given size. For each community graph $G_j$, we consider all values of $d \in [L] \setminus \{1\}$ in decreasing order, and fix the number of hyperedges of size $d$ to be 
\begin{equation}\label{eq:md}
m_d = \left\lfloor \frac {q_d}{\sum_{f=2}^{d} q_f} \left( p_j - \sum_{f = d+1}^L f m_f \right) \frac {1}{d} \right\rfloor.
\end{equation}
(We use the convention that $0/0=0$ or simply skip the values of $d$ for which $q_d=0$.) This guarantees that the distribution of hyperedge sizes follows (approximately) the desired distribution, that is, $d m_d \approx q_d p_j$. Note that, due to some possible divisibility issues, there might be some points left at the end of this process (at most $d-1$, where $d$ is the smallest value in $[L] \setminus \{1\}$ with $q_d \neq 0$, since we considered values of $d$ in decreasing order). Indeed, in order to avoid potential deficit of points, we rounded real numbers down in the definition of $m_d$ above---see equation~(\ref{eq:md}). These points are simply moved to the background graph, that is, we decrease some random values of $y_i$ (and increase the corresponding values of $z_i$); the selection of such nodes $v_i$ from community $C_j$ is made with probabilities proportional to $y_i$. As a consequence, since $p_j = \sum_{v_i \in C_j} y_i$, the value of $p_j$ is also appropriately updated, if needed, and now $p_j = \sum_{d=2}^Ld\cdot m_d$. 

Now, we need to fix the initial number of hyperedges of type $(c,d)$. (The final number of them might be slightly different.) As before, for each community graph $G_j$ and each value of $d \in [L] \setminus \{1\}$, we consider all values of $c$, $\lfloor d/2 \rfloor +1 \le c \le d$, in decreasing order, and fix the number of hyperedges of type $(c,d)$ to be
$$
m_{c,d} = \left\lfloor \frac {w_{c,d}}{\sum_{f=\lfloor d/2 \rfloor +1}^{c} w_{f,d}} \left( m_d - \sum_{f = c+1}^d m_{f,d} \right) \right\rceil.
$$

Once all $m_{c,d}$'s are computed, we know that exactly $p'_j$ points out of $p_j$ points associated with nodes in $C_j$ will belong to community hyperedges of type $(c,d)$ with majority of their neighbours from $C_j$. Clearly,
$$
p'_j = \sum_{d=2}^L \sum_{c = \lfloor d/2 \rfloor +1}^d c \cdot m_{c,d} ~~\le~~ \sum_{d=2}^L \sum_{c = \lfloor d/2 \rfloor +1}^d d \cdot m_{c,d} = \sum_{d=2}^L d\cdot m_d = p_j.
$$
Most of the remaining $p_j - p'_j$ points will typically be assigned to community hyperedges with majority of their neighbours from some other community. Those exceptional points that will end up in community hyperedges with majority of their neighbours from $C_j$ will change their type from $(c,d)$ to $(f,d)$ for some $c<f \le d$. We will explain how such assignment is done soon.

To make sure each node $v_i$ from $C_j$ is part of many community hyperedges with majority of their members from $C_j$, we further split $y_i$ points associated with the community degree of $v_i$ and assign $y'_i$ of them to be part of such community hyperedges. The value of $y'_i \le y_i$ that we aim to be close to $y_i\cdot p'_j/p_j$ is chosen as follows. To guarantee that each $y'_i$ is at least $\lfloor y_i\cdot p'_j/p_j \rfloor$ we initially set this value for every $y'_i$. Next, we compute by how much the sum of such floors is less than $p'_j$, that is, we set $t := p'_j - \sum_{v_i \in C_j} \lfloor y_i\cdot p'_j/p_j \rfloor$. Clearly, $0 \le t < |C_j|$. Then, we randomly sample, without replacement, $t$ points (with probability proportional to $y_i\cdot p'_j/p_j - \lfloor y_i\cdot p'_j/p_j \rfloor$). Such points have their corresponding values of $y'_i$ increased by one that is, $y'_i = \lceil y_i\cdot p'_j/p_j \rceil$. As a result, $\sum_{v_i\in C_j}y'_i=p'_j$ and no value of $y'_i$ is more than 1 away from its desired value of $y_i\cdot p'_j/p_j$.

\medskip

Let us summarize properties of the auxiliary partition of points that we just created as it is crucial for the process of creating the \hABCD\ hypergraph. There are $\sum_{i \in [n]} x_i = \vol(V)$ points in total; $p = \sum_{i \in [n]} z_i$ of them are put aside to form a set $\mathcal{B}$ that will be used to create the background graph. From $p_j = \sum_{v_i \in C_j} y_i$ remaining points associated with nodes in $C_j$ we selected $p'_j = \sum_{v_i \in C_j} y'_i$ points to form a set $\mathcal{C}_j$ that will be used to create community hyperedges with majority of members from $C_j$. The remaining $\sum_{i \in [n]} (y_i - y'_i) = \sum_{j=1}^{\ell} (p_j - p'_j)$ points form a set $\mathcal{C}$.

Finally, we are ready to create random community graphs $G_j$, $j \in [\ell]$. For each community hyperedge of type $(c,d)$ that belongs to community $C_j$, we randomly select $c$ points from $\mathcal{C}_j$ (without replacement). At this point all points from $\mathcal{C}_j$ are  exhausted. Once we process all community hyperedges of all community graphs, points from $\mathcal{C}$ are randomly assigned to community hyperedges so that their sizes match the desired values, that is, a hyperedge of type $(c,d)$ gets additional $d-c$ points. As mentioned earlier, note that these points are typically taken from outside of the community a given hyperedge of type $(c,d)$ belongs to, but it might happen that some hyperedge becomes of type $(f,d)$ for some value of $f$, $c < f \le d$. 

\subsubsection*{Creating Background Graph}

Background graph is created similarly to community graphs. There are $p = \sum_{i \in [n]} z_i$ points in set $\mathcal{B}$ that will be associated with background hyperedges of $G_0$. As before, we consider all values of $d \in [L] \setminus \{1\}$ in decreasing order, and fix $m_d$, the number of hyperedges of size $d$ in the background graph, using formula~(\ref{eq:md}).

We first randomly partition points in $\mathcal{B}$ into sets of appropriate sizes that become background hyperedges. Note that after this process, there could be $r \ge 0$ points left, where $0 \le r < R = \min\{ d \in [L] \setminus \{1\} : q_d > 0\}$. (In particular, when edges are allowed, $R=2$ and so there is at most one point left.). If $r=0$, then we are done, and so in what follows we discuss how the case $r>0$ is handled.

First we check whether $q_1>0$. If this is the case, then for the multi-hypergraph variant of the model, we simply create $r$ additional hyperedges of size 1 from the remaining $r$ points. On the other hand, if simple hypergraph is requested, then we check whether all remaining points are assigned to unique nodes and none of them corresponds to an already created hyperedge of size 1. If both of these constraints are satisfied, then one more time we create $r$ hyperedges of size 1 from these points.

If $q_1=0$, or $q_1>0$ and simple hypergraphs were requested but at least one of the above constraints were not satisfied, then we increase the background degrees $z_i$ of $R - r$ random nodes making room for one additional hyperedge of size $R$ (that is, we increase $m_R$ by 1); the selection of such nodes $v_i$ is made with probabilities proportional to $z_i$.

Finally, let us mention that background hyperedges might produce hyperedges of type $(c,d)$ for some $d/2 < c \le d$. However, for large graphs with many communities there will typically not be very many of them. Type $(2,3)$ has the best change to get additional ``boost'', followed by type $(2,2)$.

\subsubsection*{Creating Simple Hypergraphs---Rewiring}

In the variant of the model that produces simple hypergraphs, it remains to deal with multi-sets or hyperedge repetitions that might potentially get created.

We take all hyperedges generated so far and split them into two objects. We put all hyperedges that are sets into set $S$.  All the remaining hyperedges (that is, hyperedges that are multi-sets or hyperedges that were duplicated) are put into vector $B$. Note that in case of having $t \ge 2$ duplicates of a given hyperedge that is a set, one such hyperedge goes to $S$ and $t-1$ of them go to $B$. Next, we try to fix problematic hyperedges from vector $B$ as follows. In one round, we pick a hyperedge $b$ from vector $B$. We compute the \emph{indisposition} of $b$ as the sum of two terms. The first term is equal to 1 if $b$ is in $S$, and 0 otherwise. The second term is equal to the number of duplicates in $b$. Note that exactly one element of this sum is positive.

Next, we randomly pick a good hyperedge $g$ from $S$. Let $s_b$ be the size of $b$ and $s_g$ be the size of $g$. We merge points from $b$ and $g$ and randomly split them into new hyperedges, $h_1$ and $h_2$, of sizes $s_b$ and $s_g$. Now, we calculate the sum of indispositions of $h_1$ and $h_2$ the same way as it was done for $b$. If the total indispositions is less than the initial indisposition of $b$, then we remove $b$ and $g$ from $S$ and, respectively, $B$, and put $h_1$ and $h_2$ back into these sets (to $S$ if they have indispositions of zero, and to $B$ otherwise). This process is repeated no more than $\texttt{maxiter} \cdot|B|$ times (that is, \texttt{maxiter} times per one initial bad hyperedge); in implementation, $\texttt{maxiter}=100$. If after that many iterations there are still some bad edges left in $B$, then we give up and the generation process is terminated and appropriate warning is returned.

\subsection{Distribution of Hyperedges Sizes}

Clearly, modelling complex networks as hypergraphs is still in an early stage but the initial experiments, such as the ones done on 13 real-world hypergraphs (publication coauthors, drug abuse warning network drugs, emails from an European research institution, national drug code directory drug, online question tags, and thread participants) suggest that in many networks the largest hyperedges are typically of a small size, not comparable to the number of nodes of the hypergraph~\cite{do2020structural}. As a result, parameter $L$ in \hABCD\ should be set to be some relatively small value that does not grow with the order of the network.

Having said that, the datasets used in~\cite{do2020structural} are part of the Benson's collection\footnote{\url{https://www.cs.cornell.edu/~arb/data/}}, which contains real-world hypergraphs with a wide range of different maximum hyperedge sizes. There are datasets with large values of $L$ such as the network of Amazon reviews with $n = 2{,}268{,}231$ and $L = 9{,}350$ or the network of stackoverflow answers with $n = 15{,}211{,}989$ and $L = 61{,}315$.

\medskip

In this paper, since we aim for a synthetic model for community detection, we concentrate on modelling networks with relatively small hyperedges. Indeed, enormous hyperedges are often regarded as noise in this context and removed during the preprocessing phase. Because of that, our goal is rather ambitious and we aim to generate hypergraphs with a distribution of hyperedges as uniform as possible. Insisting on this important property is inherently computationally expensive for large values of $L$ (see Section~\ref{sec:time_complexity}). Dealing with larger values of $L$ needs a slightly different approach but is doable after scarifying uniformity condition (see Conclusions, Section~\ref{sec:future}).

\section{Experiments}\label{sec:experiments}

In this section we present a series of experiments investigating various properties of the \hABCD\ model. For such experiments, unless otherwise indicated, we used the following parameters: $\gamma = 2.5$, $\delta = 5$, $D = n^{\zeta}$ with $\zeta = 0.5$ (parameters affecting the degree distribution),  $\beta = 1.5$, $s = 50$, $S=n^\tau$ with $\tau = 3/4$ (parameters affecting the distribution of community sizes),  $\xi=0.2$ (the level of noise), and  $q_2 = q_3 = q_4 = q_5 = 0.25$ (the distribution of hyperedge sizes). In the first few experiments, community edges could have any number of nodes from its own community as long as majority of them belong to that community. In other words, for a fixed value of $d$, we consider a uniform distribution of $w_{c,d}$. As mentioned at the very beginning of this section, we refer to this distribution as majority model. However, other models will be investigated later on. In all cases, we generate simple hypergraphs. 

\subsection{Degree Distribution}

The degree distribution, by design, follows power-law with exponent $\gamma$ and from that perspective there is no difference between \hABCD\ and the original model for graphs, \ABCD. As a result, we have a good understanding of its asymptotic behaviour~\cite{kaminski2022modularity}. We start with a simple experiment to see whether theoretical, asymptotic results can be used to predict the empirical behaviour for relatively small values of $n$.  

\begin{figure}[h]
     \centering
     \includegraphics[width=0.45\textwidth]{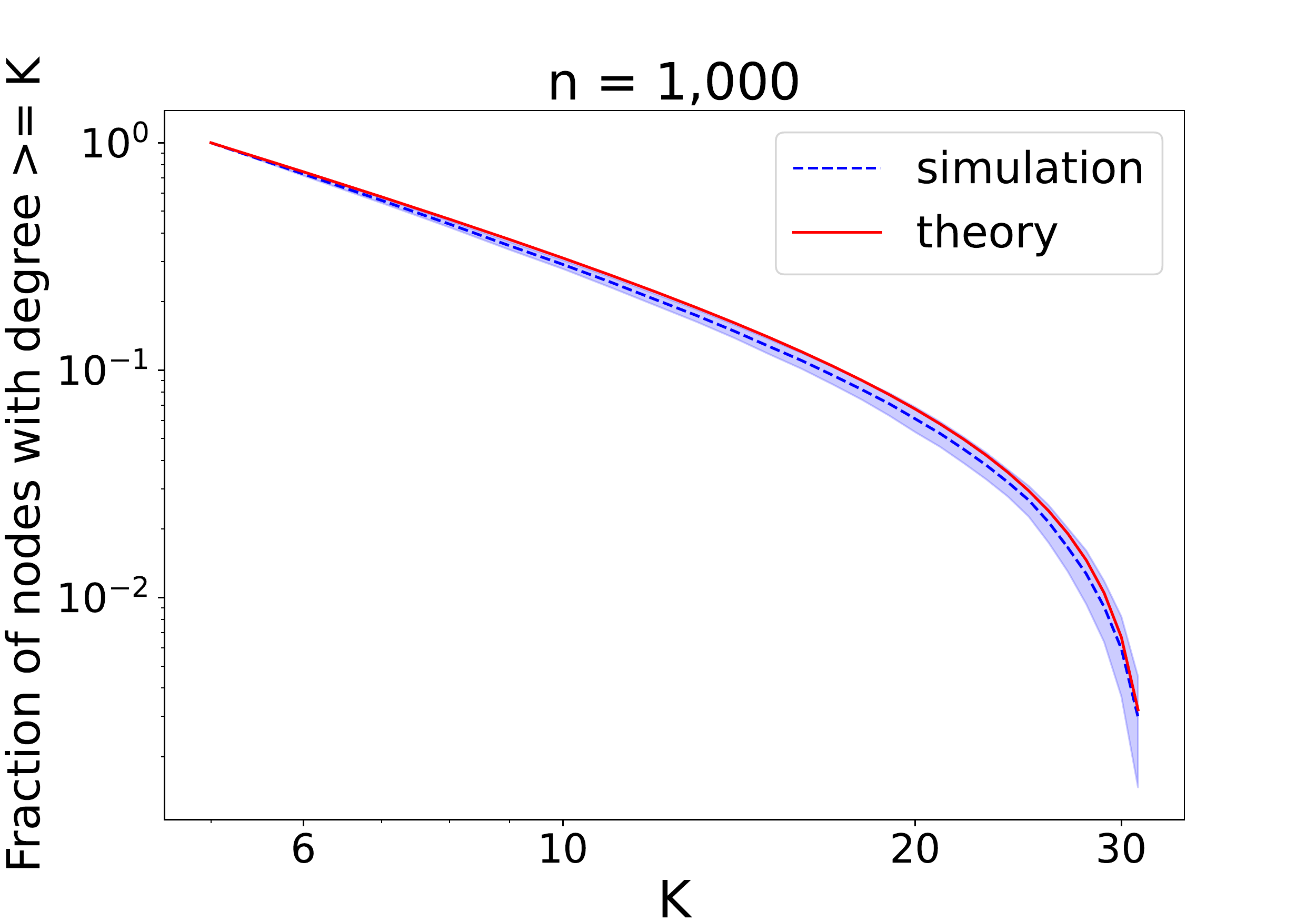}
     \includegraphics[width=0.45\textwidth]{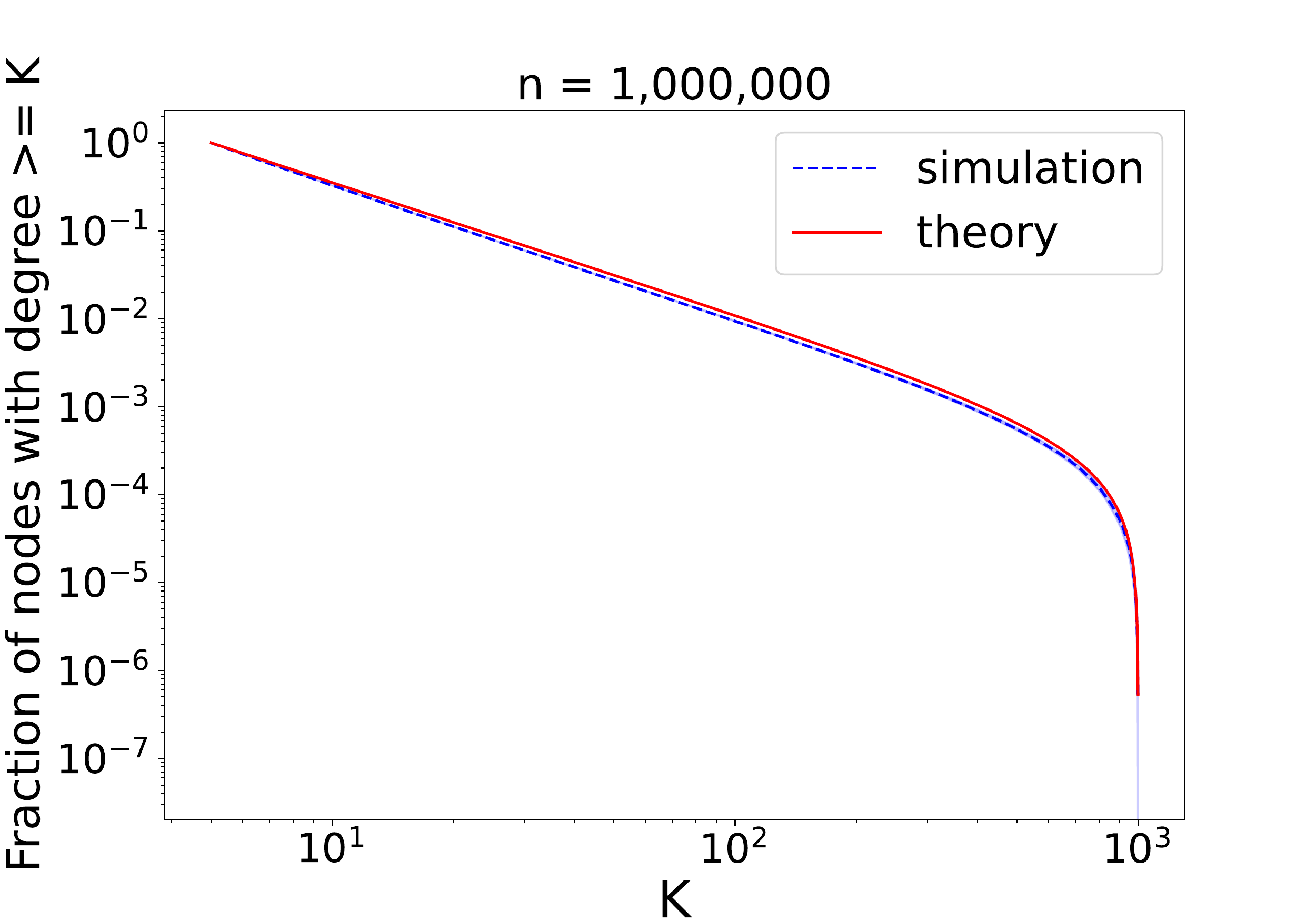}
     \caption{Comparison of the degree distribution predicted by theoretical, asymptotic results with results from simulations.}
     \label{fig:degree}
\end{figure}

In Figure~\ref{fig:degree}, we plot the fractions of nodes with degree larger than or equal to $K$ for all $K \in \{\delta, \delta+1, \ldots, D\}$. Results are presented for small ($n=1,000$) and larger ($n=1,000,000$) graphs. For each graph size, 100 independent runs were performed and the shaded areas correspond to the standard deviation for each value. We compare those simulation results with the values predicted by theory. We observe a good correspondence even for small graphs, with almost perfect match for larger graphs. Additionally in Table~\ref{tab:ccount}, as a reference, we give mean and standard deviation of the number of communities for $n=2^i$, $i \in \{10, \ldots, 20\}$.

\begin{table}
\centering
\begin{tabular}{r|rr}
\toprule
\multicolumn{1}{c|}{$n$} & mean $\ell$ & std $\ell$ \\
\midrule
    1,024 &  11.03 &  1.29 \\
    2,048 &  16.81 &  2.23 \\
    4,096 &  26.61 &  3.52 \\
    8,192 &  39.41 &  6.01 \\
   16,384 &  62.12 &  8.39 \\
   32,768 &  96.60 & 14.92 \\
   65,536 & 148.49 & 21.74 \\
  131,072 & 229.72 & 28.61 \\
  262,144 & 352.59 & 40.51 \\
  524,288 & 539.00 & 54.11 \\
1,048,576 & 844.01 & 82.39 \\
\bottomrule
\end{tabular}
\caption{Mean and standard deviation of the number of communities $\ell$ as a function of $n$. Reported values are computed using 100 independent samples.\label{tab:ccount}}
\end{table}

\subsection{Distribution of Community Sizes}

The process of generating community sizes in \hABCD\ is the same as in \ABCD, and it is designed to follow (truncated) power-law distribution with exponent $\beta$. Asymptotic behaviour is then quite easy to analyze~\cite{kaminski2022modularity}. However, since there are substantially less communities than nodes, one should not expect to have equally good behaviour of the distribution of community sizes as for the degree distribution. 

\begin{figure}[h]
     \centering
     \includegraphics[width=0.45\textwidth]{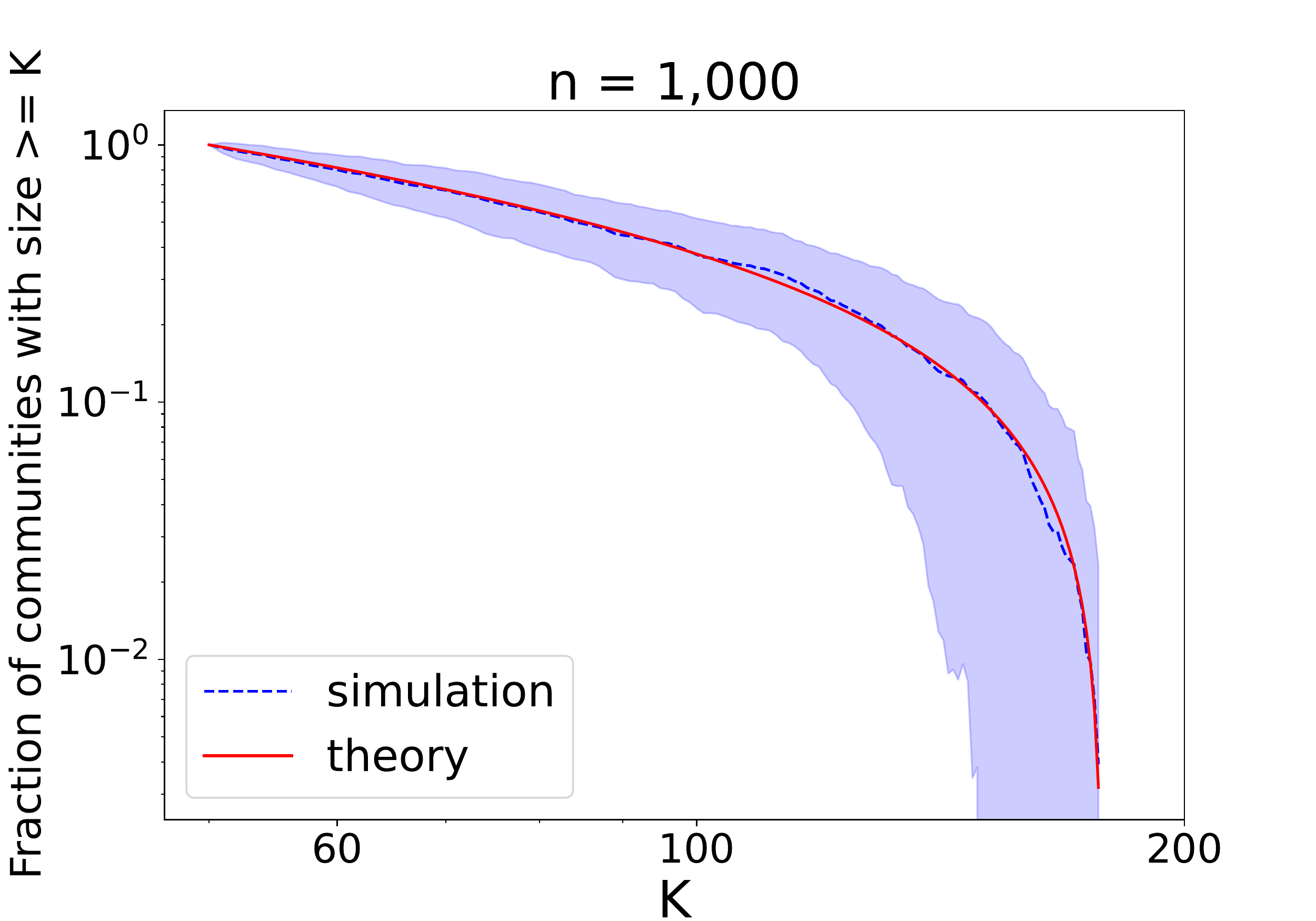}
     \includegraphics[width=0.45\textwidth]{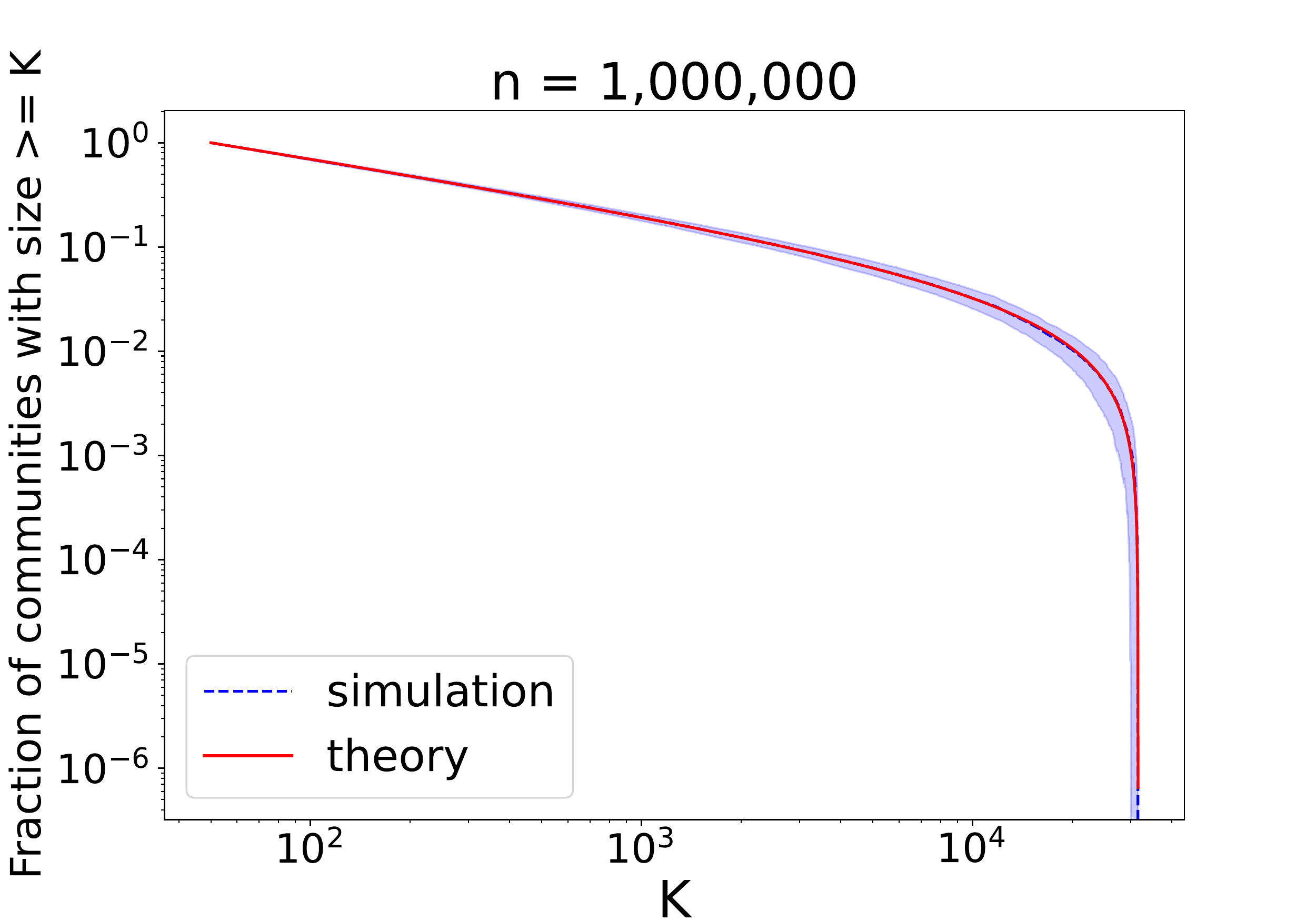}
     \caption{Comparison of the distribution of community sizes predicted by theoretical, asymptotic results with results from simulations.}
     \label{fig:commsize}
\end{figure}

In Figure~\ref{fig:commsize}, we plot the fractions of communities with sizes larger than or equal to $K$ for all $K \in \{s, s+1, \ldots, S\}$. As in the previous experiments, results are presented for small ($n=1,000$) and larger ($n=1,000,000$) graphs. For each graph size, we performed 100 independent runs and compared simulation results with the values predicted by theory. The conclusions are similar but, as expected and mentioned earlier, due to the fact that the number of communities is much smaller than the number of nodes, the standard deviations are substantially larger than those for the degree distribution presented in Figure~\ref{fig:degree}.

\subsection{Distribution of Edge Sizes}

The distribution of edge sizes is controlled by vector $q$. In our experiment, we requested that the same fraction of the total volume to be associated with edges of sizes between 2 and 5 ($q_2=q_3=q_4=q_5=0.25$). However, because of rounding (see~(\ref{eq:md})), there is a mild bias towards smaller hyperedges. For large hypergraphs, the difference should not be visible but for small ones could be. The goal of the next experiment is to investigate how strong the bias is.

We independently generated 100 graphs for each size $n=2^i$, $i \in \{10, \ldots, 20\}$. We considered two different values for the parameter responsible for the level of noise, respectively $\xi=0.2$ and $\xi=0.7$. The choice of hyperedge composition (matrix $w$) does not affect the distribution of edge sizes so we select the majority model, one of the three standard models of \hABCD.

\begin{figure}[h]
     \centering
     \includegraphics[width=0.9\textwidth]{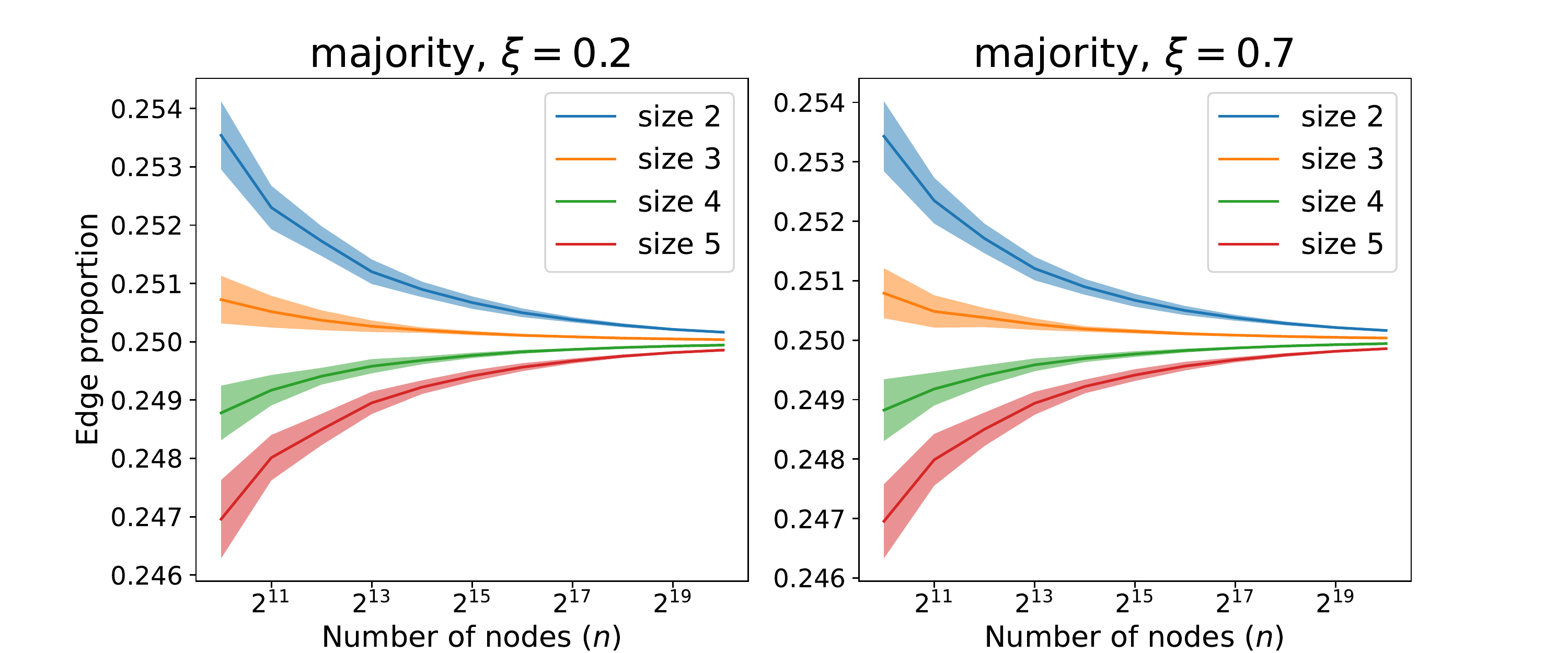}
     \caption{Distribution of the total volume associated with hyperedge sizes, given parameters $q_2 = q_3 = q_4 = q_5 = 0.25$. The majority model and two values of the parameter responsible for the level of noise ($\xi = 0.2, 0.7$) were used.}
     \label{fig:q_dist}
\end{figure}

The results of the experiment are shown in Figure~\ref{fig:q_dist}. As expected, we see a small bias toward smaller edges for small graphs, and convergence toward a uniform distribution for larger graphs. However, the difference, even for the smallest graphs on $2^{10}$ nodes, is quite small. Parameter $\xi$ might potentially affect the distribution. To see it, consider hyperedges of the largest size $d$. For a small value of parameter $\xi$, almost all communities (if not all) have their volumes much larger than $d$. Such communities are expected to assign, on average, $(d-1)/2$ less points to the largest hyperedges than requested. For small graphs, the number of communities is not negligible in comparison to the total number of edges, and so we expect to see the difference. On the other hand, if $\xi$ is very close to 1, most of the volume is assigned to the background graph which may affect at most one edge. Hence, the bias should be small for such cases, even for small graphs (of course, large values of $\xi$ are not of practical importance). In our experiments, the results for the two values of $\xi$ are indistinguishable; as expected, $\xi$ affects which nodes are put into hyperedges but not the distribution of hyperedge sizes.

Additionally in Table~\ref{tab:hecount}, as a reference, we give mean and standard deviation of the number of generated hyperedges $m$ for $n=2^i$, $i \in \{10, \ldots, 20\}$, majority model and $\xi=0.2$ (type of the model and $\xi$ would not affect the results). Note that the number of hyperedges is well concentrated around the mean.

\begin{table}
\centering
\begin{tabular}{r|rr}
\toprule
\multicolumn{1}{c|}{$n$} & mean $m$ & std $m$ \\
\midrule
    1,024 &     2,969 &   48.11 \\
    2,048 &     6,325 &   92.79 \\
    4,096 &    13,369 &  185.96 \\
    8,192 &    28,056 &  311.11 \\
   16,384 &    58,408 &  492.11 \\
   32,768 &   120,715 &  956.52 \\
   65,536 &   248,123 & 1386.05 \\
  131,072 &   508,032 & 2,201.55 \\
  262,144 & 1,035,753 & 3,631.05 \\
  524,288 & 2,105,073 & 5,278.08 \\
1,048,576 & 4,267,800 & 7,893.15 \\
\bottomrule
\end{tabular}
\caption{Mean and standard deviation of the number of hyperedges $m$ as a function of $n$. Reported values are computed using 100 independent samples.\label{tab:hecount}}
\end{table}

\subsection{Distribution of Hyperedge Composition}

The setup for this experiment is exactly the same as in the previous subsection, but this time we compare the distribution of hyperedge composition, that is, the number of hyperedges of type $(c,d)$ for $d \in \{2,3,4,5\}$ and $d/2 < c \le d$. We say that the remaining hyperedges are of type $(0,d)$, that is, hyperedges of type $(0,d)$ are hyperedges of size $d$ that do not have a majority of nodes in any of the communities.

\begin{figure}[h]
     \centering
     \includegraphics[width=0.6\textwidth]{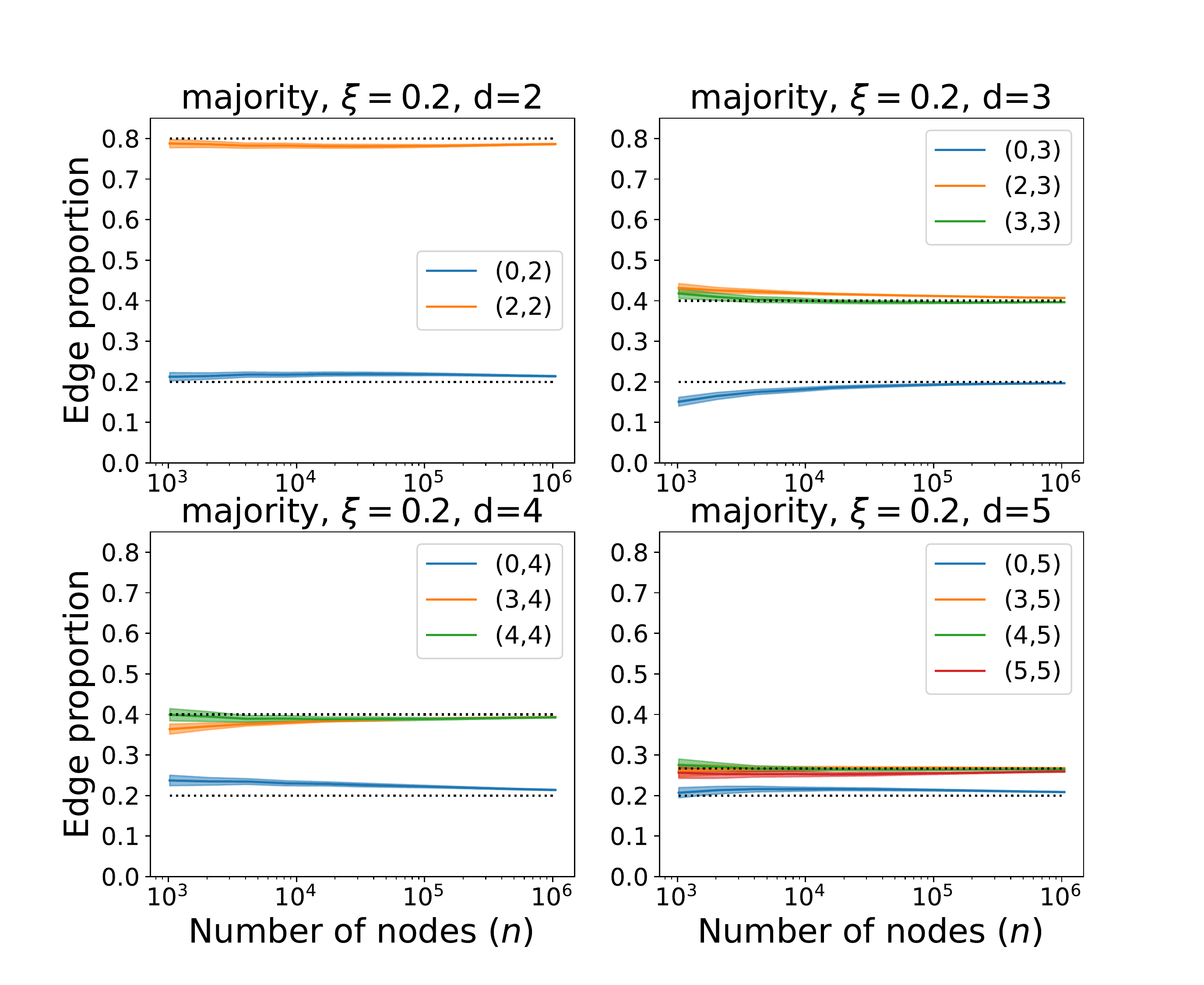}
     \includegraphics[width=0.6\textwidth]{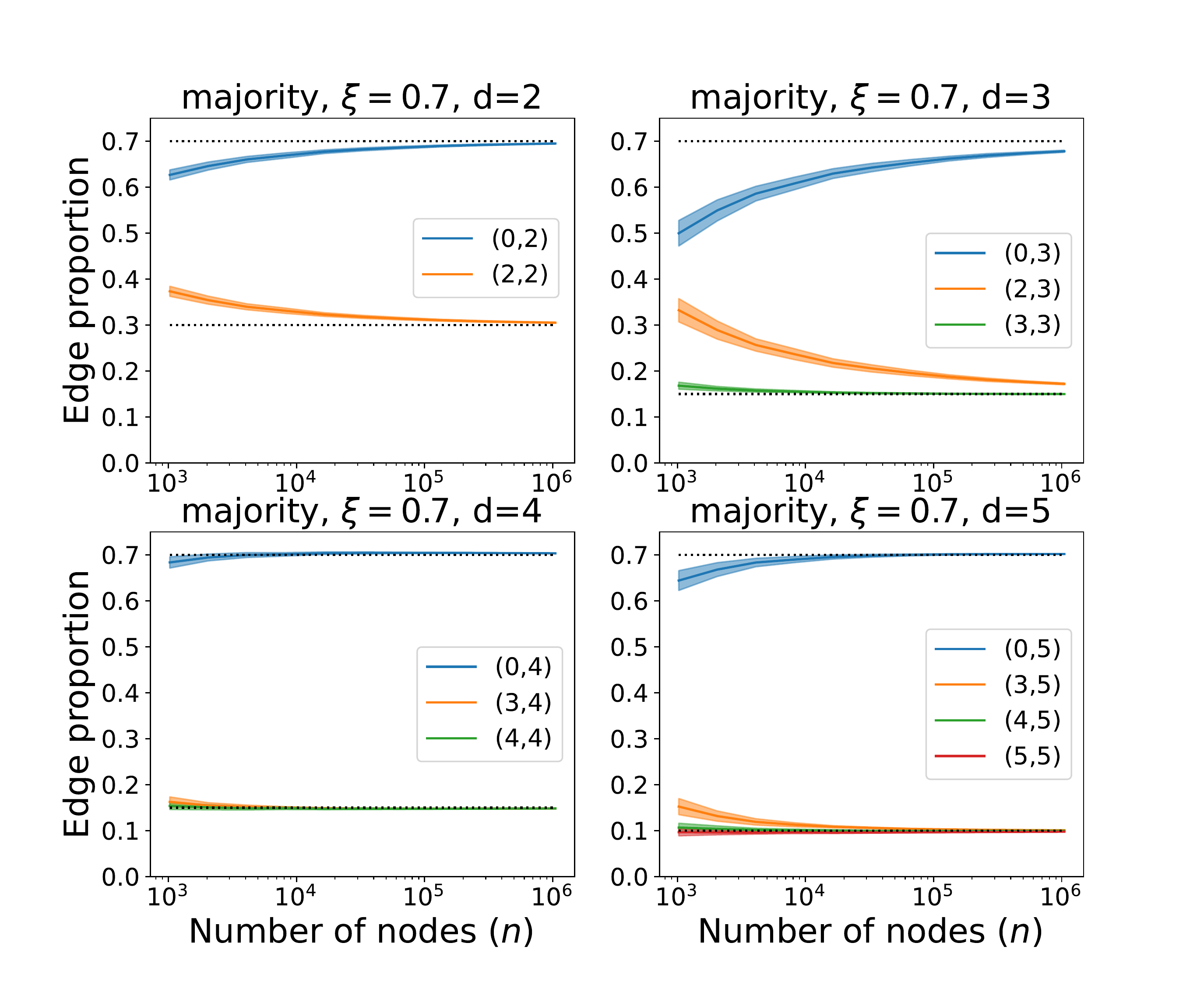}
     \caption{Distribution of type $(c,d)$ hyperedges for $d \in \{2,3,4,5\}$: {\em majority} model with $\xi=0.2$ (top 4 figures) and $\xi=0.7$ (bottom 4 figures).}
     \label{fig:wcd_1}
\end{figure}

\begin{figure}[h]
     \centering
     \includegraphics[width=0.6\textwidth]{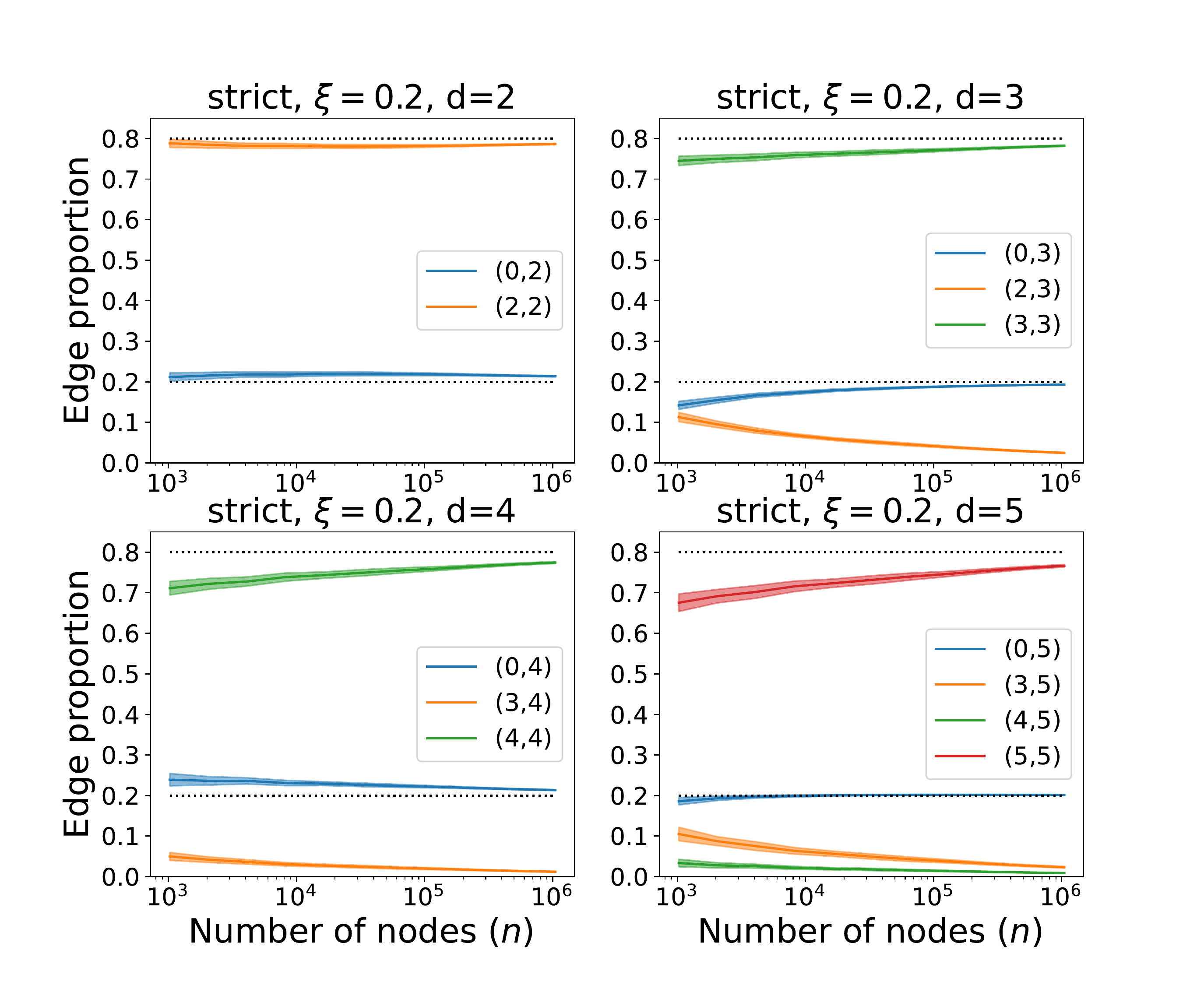}
     \includegraphics[width=0.6\textwidth]{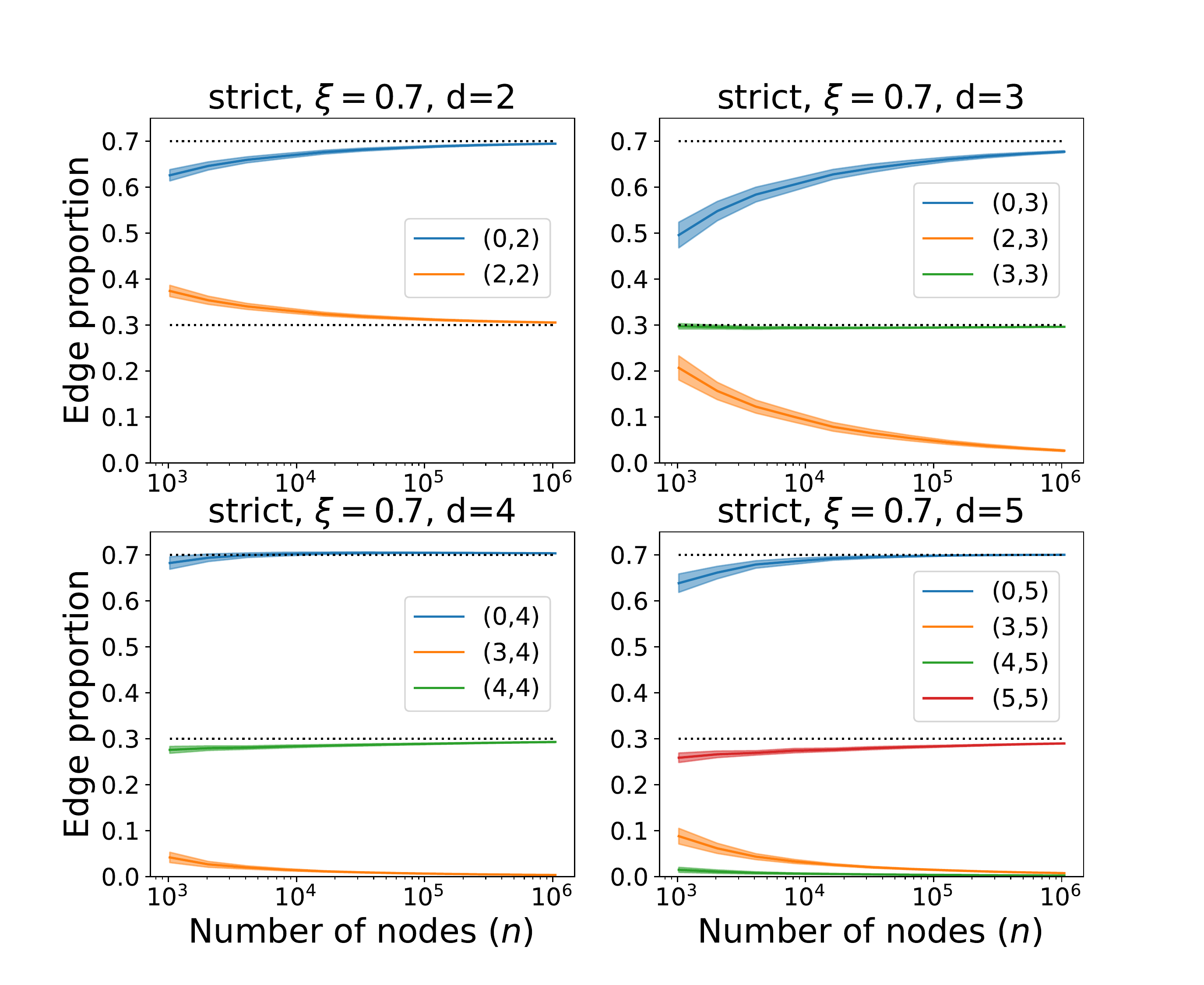}
     \caption{Distribution of type $(c,d)$ hyperedges for $d \in \{2,3,4,5\}$: {\em strict} model with $\xi=0.2$ (top 4 figures) and $\xi=0.7$ (bottom 4 figures).}     
     \label{fig:wcd_3}
\end{figure}

Recall that the initial number of hyperedges of type $(c,d)$ is governed by parameter $w_{c,d}$ stored in matrix $w$. However, some hyperedges of type $(c,d)$ might ``get promoted'' and become of type $(f,d)$ for some $c < f \le d$, if at least one of the additional $d-c$ points are taken from the same community. Moreover, some of the background hyperedges might become of type $(c,d)$ for some $d/2 < c \le d$. Finally, in the experiments we produce simple hypergraphs (recall that this is a default option but the framework also allows for generation of multi-hypergraphs). The requirement of generating simple hypergraphs means that when a multi-hyperedge or parallel hyperedge is generated, then it has to be rewired as discussed in Section~\ref{sec:model}. The rewiring process, if needed, performed on an edge of type $(c, d)$ typically reduces its value of $c$. This is especially visible for hyperedges of type $(d,d)$ that are most likely to be multi-hyperedges for large $d$ and small communities. For large graphs, the discussed discrepancies should not be noticeable but small graphs might have distribution of hyperedges types slightly different than the desired one, requested via matrix $w$.  The results for two standard models (majority and strict) and $\xi \in \{0.2, 0.7\}$ are shown in Figure~\ref{fig:wcd_1} (majority) and Figure~\ref{fig:wcd_3} (strict).

%

%

There are several interesting observations that can be made from those plots. Let us first start from the majority model (Figure~\ref{fig:wcd_1}). The requested fraction of hyperedges of size $d$ that are of type $(0,d)$ is $\xi$, and those that are of type $(c,d)$ for some $d/2 < c \le d$ should have frequency close to $(1-\xi)w_{c,d} = (1-\xi)/\lceil d/2 \rceil$, regardless of the value of $c$. Large hypergraphs, as expected, have distributions that are very close to the requested one. For smaller graphs we see that some of the background edges got ``promoted'' and became of type $(c,d)$ for some $d/2 < c \le d$. As a result, functions associated with types $(0,d)$ for $\xi = 0.7$ tend to increase with the size of the hypergraph. On the other hand, for $\xi=0.2$ and, say, $d=5$ the function associated with type $(0,5)$ seems to be initially larger than $0.2$ before converging to $0.2$. This is because the total number of community hyperedges of size $d=5$ is smaller than requested (as discussed in the previous subsection) and so background hyperedges consume a larger share of the total number of hyperedges of size $d=5$. We also notice that some types of hyperedges are more likely than others, despite the fact that their corresponding values of $w_{c,d}$ are equal. For example, with $d=3$, we see more hyperedges of type $(2,3)$ than of type $(3,3)$ (larger deviation for $\xi=0.7$ than for $\xi=0.2$).  This is, again, due to hyperedges from the background graph that are more likely to generate such edges by pure randomness. 

Similar observations can be made for the {\em strict} model (Figure~\ref{fig:wcd_3}). As in the previous situation, the requested fraction of hyperedges of size $d$ that are of type $(0,d)$ is $\xi$. In fact, the initial distribution of hyperedges of type $(0,d)$ is exactly the same as last time (before rewiring), as the processes of generating background graphs are the same. However, this time the bias introduced by rewiring is slightly larger than for the majority rule, as hyperedges of type $(d,d)$ are most likely to be non-simple. This discrepancy diminishes as size of the graph increases, as then the fraction of hyperedges that need to be rewired drops. The fraction of hyperedges that are of type $(d,d)$ is requested to be $(1-\xi)w_{c,d}=1-\xi$ but other types should not be present at all. The distribution of hyperedges types is very close to the desired ones for large graphs. For small graphs, while most hyperedges are either of type $(d,d)$ or type $(0,d)$, we see some of them with $d/2<c<d$, again, a side effect of the background graph. Not surprisingly, there are more of type $(3,5)$ than of type $(4,5)$. 

\subsection{Modularity Function and the Need for Matrix $w$}

As mentioned in the introduction, there are many ways members of one community could form hyperedges. In some real-world networks, most hyperedges are homogeneous (recall an example with papers written by mathematicians) but in some other networks many hyperedges are heterogeneous (again, recall an example with papers written by medial doctors). Hence, \hABCD\ model has to be able to simulate various scenarios. The main reason behind experiments in this subsection is to show that one may use \hABCD\ model to generate hypergraphs that are indistinguishable from their 2-section point of view but, at the same time, have completely different structures when viewed as hypergraphs. 

\medskip

Before we move to our experiments, we need to introduce a few definitions. The modularity function for graphs was first introduced by Newman and Girvan in~\cite{newman2004finding} and is currently often used to measure the presence of community structure in networks. Many popular algorithms for partitioning nodes of large graphs use it~\cite{clauset2004finding,lancichinetti2011limits,newman2004fast} and perform very well. The modularity function favours partitions of the set of nodes of a graph in which a large proportion of the edges fall entirely within the parts, but benchmarks it against the expected number of edges one would see in those parts in the corresponding Chung-Lu random graph model. 

Formally, for a graph $G=(V,E)$ and a given partition $\A = \{A_1, A_2, \ldots, A_k\}$ of $V$, the \emph{modularity function} is defined as follows:
\begin{eqnarray}
q_G(\A) &=& \sum_{A_i \in \A} \frac{e_G(A_i)}{|E|}  - \sum_{A_i \in \A} \left( \frac{\vol_G(A_i)}{\vol_G(V)} \right)^2, \label{eq:q_G_A}
\end{eqnarray}
where $e_G(A_i) = |\{ \{v_j,v_k\} \in E : v_j, v_k \in A_i\}|$ is the number of edges in the subgraph of $G$ \emph{induced by} set $A_i$ and $\vol_G(A_i) = \sum_{v_j \in A_i} \deg_G(v_j)$. The first term in~(\ref{eq:q_G_A}), $\sum_{A_i \in \A} e_G(A_i)/|E|$, is called the \emph{edge contribution} and it computes the fraction of edges that fall within one of the parts. The second one, $\sum_{A_i \in \A} (\vol_G(A_i)/\vol_G(V))^2$, is called the \emph{degree tax} and it computes the expected fraction of edges that do the same in the corresponding random graph (the null model). The modularity measures the deviation between the two. The maximum \emph{modularity} $q^*(G)$ is defined as the maximum of $q_G(\A)$ over all possible partitions $\A$ of $V$; that is, $q^*(G) = \max_{\A} q_G(\A).$

For edges of size greater than 2, several definitions can be used to quantify the edge contribution for a given partition $\A$ of the set of nodes. As a result, the choice of hypergraph modularity function is not unique; it depends on how strongly one believes that a hyperedge is an indicator that some of its nodes fall into one community. The fraction of nodes of a given hyperedge that belong to one community is called its \emph{homogeneity} and it is assumed that it is more than 50\%. As a result, we are guaranteed that hyperedges contribute to at most one part. Once a concrete variant is fixed, one needs to benchmark the corresponding edge contribution using the degree tax computed for the generalization of the Chung-Lu model to hypergraphs proposed in~\cite{kaminski2019clustering}.

In~\cite{kaminski2020community}, various definitions of modularity functions from~\cite{kaminski2019clustering} were put into a common framework, and are available in the {\tt HyperNetX} library\footnote{\url{https://github.com/pnnl/HyperNetX}}. This general framework is flexible and so can be tuned and applied to hypergraphs with hyperedges of different homogeneity. For each hyperedge size $d$, we will independently deal with contribution to the modularity function coming from hyperedges of size $d$ with precisely $c$ members from one of the parts, where $c > d/2$. For $d \in \N$ and $p \in [0,1]$, let $\textrm{Bin}(d,p)$ denotes the binomial random variable with parameters $d$ and $p$. Let 
$$
q_{H}^{c,d}({\mathbf A}) = \frac{1}{|E|} \sum_{A_i \in {\bf A}} 
\left( e_H^{d,c}(A_i) - |E_d| \cdot \textrm{P} \left( \textrm{Bin} \left( d, \frac{\vol(A_i)}{\vol(V)} \right) = c \right) \right),
$$
where $e_H^{d,c}(A_i)$ is the number of hyperedges of size $d$ that have exactly $c$ members in $A_i$. The hypergraph modularity function is controlled by \emph{hyper-parameters} $u_{c,d} \in [0,1]$ ($d \ge 2$, $\lfloor d/2 \rfloor + 1 \le c \le d$). For a fixed set of hyper-parameters, we simply define
\begin{equation}\label{eq:new_modularity}
q_{H}({\mathbf A}) = \sum_{d \geq 2} \sum_{c = \lfloor d/2 \rfloor + 1}^{d} u_{c,d} \ q_{H}^{c,d}({\mathbf A}).
\end{equation}
This definition gives us a lot of flexibility and allows us to value hyperedges of some types $(c,d)$ more than others, depending on their size and level of homogeneity. The choice of these hyper-parameters depends on how strongly we believe that a hyperedge is an indicator that nodes belonging to it fall into one community. In our experiments, we restricted ourselves to three families of parameters $u_{c,d}$, corresponding to the three standard models of $w_{c,d}$ used in the \hABCD\ model: majority, linear, and strict.

\medskip

\begin{figure}[h]
     \centering
     \includegraphics[width=0.45\textwidth]{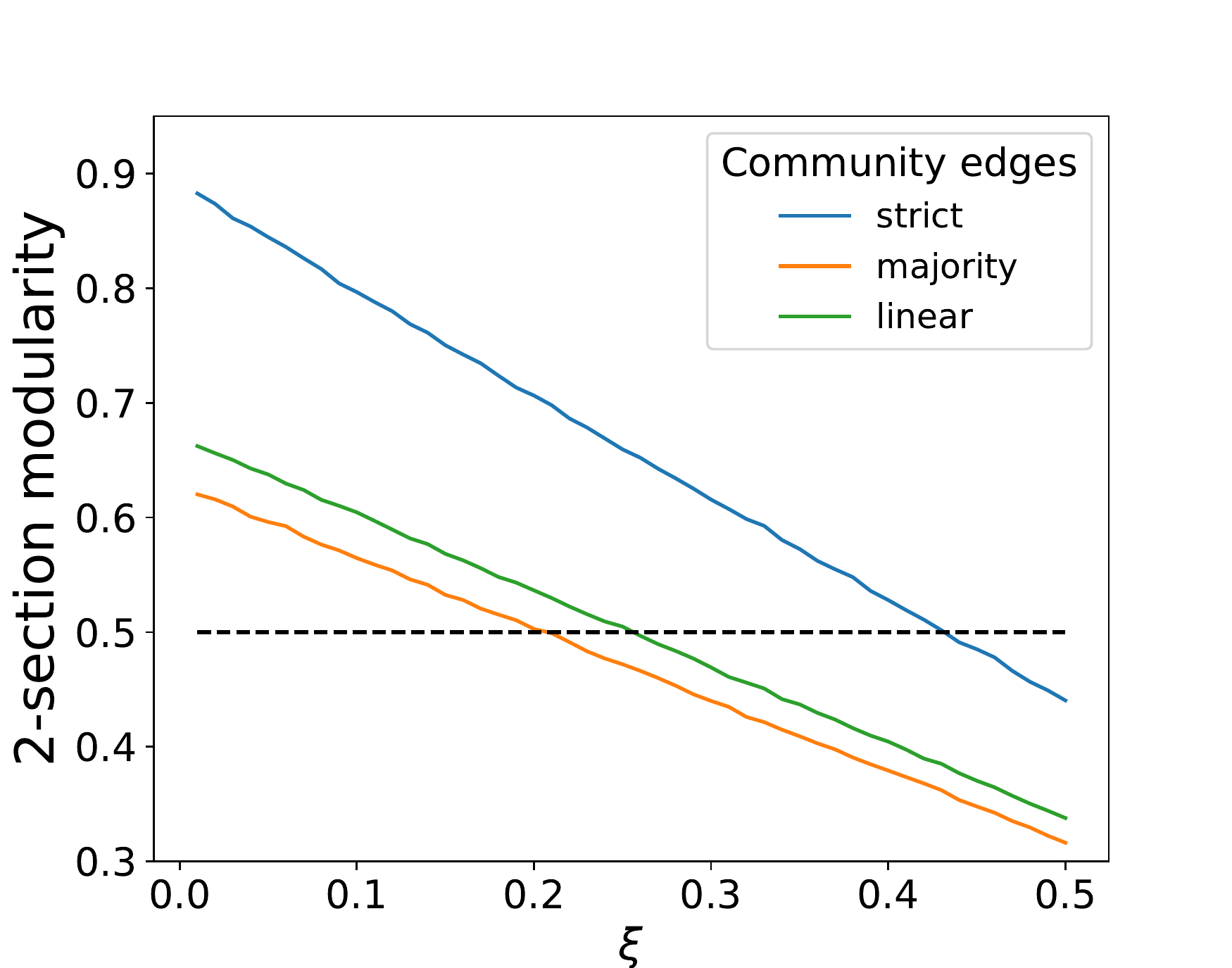}
     \includegraphics[width=0.45\textwidth]{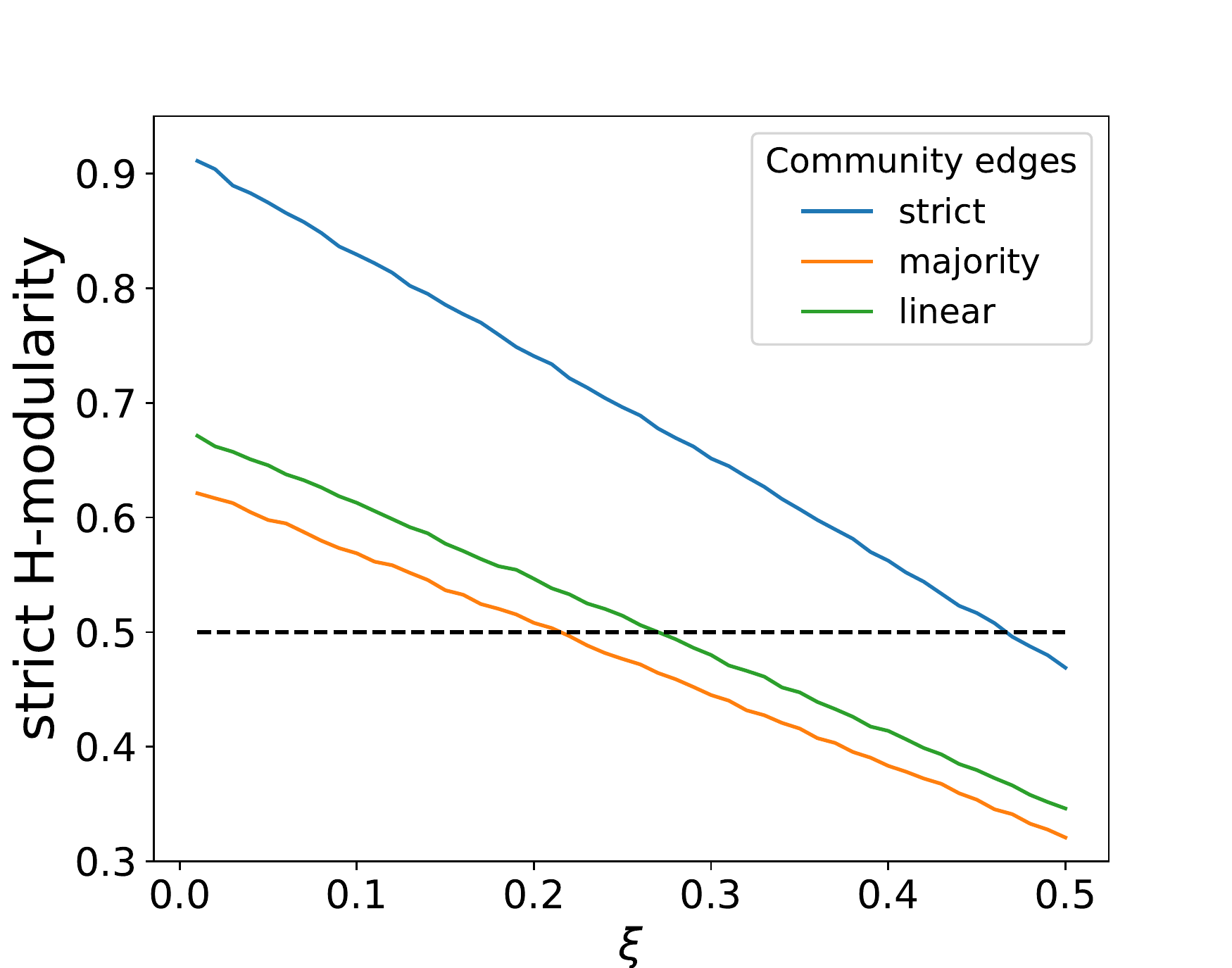}
     \includegraphics[width=0.45\textwidth]{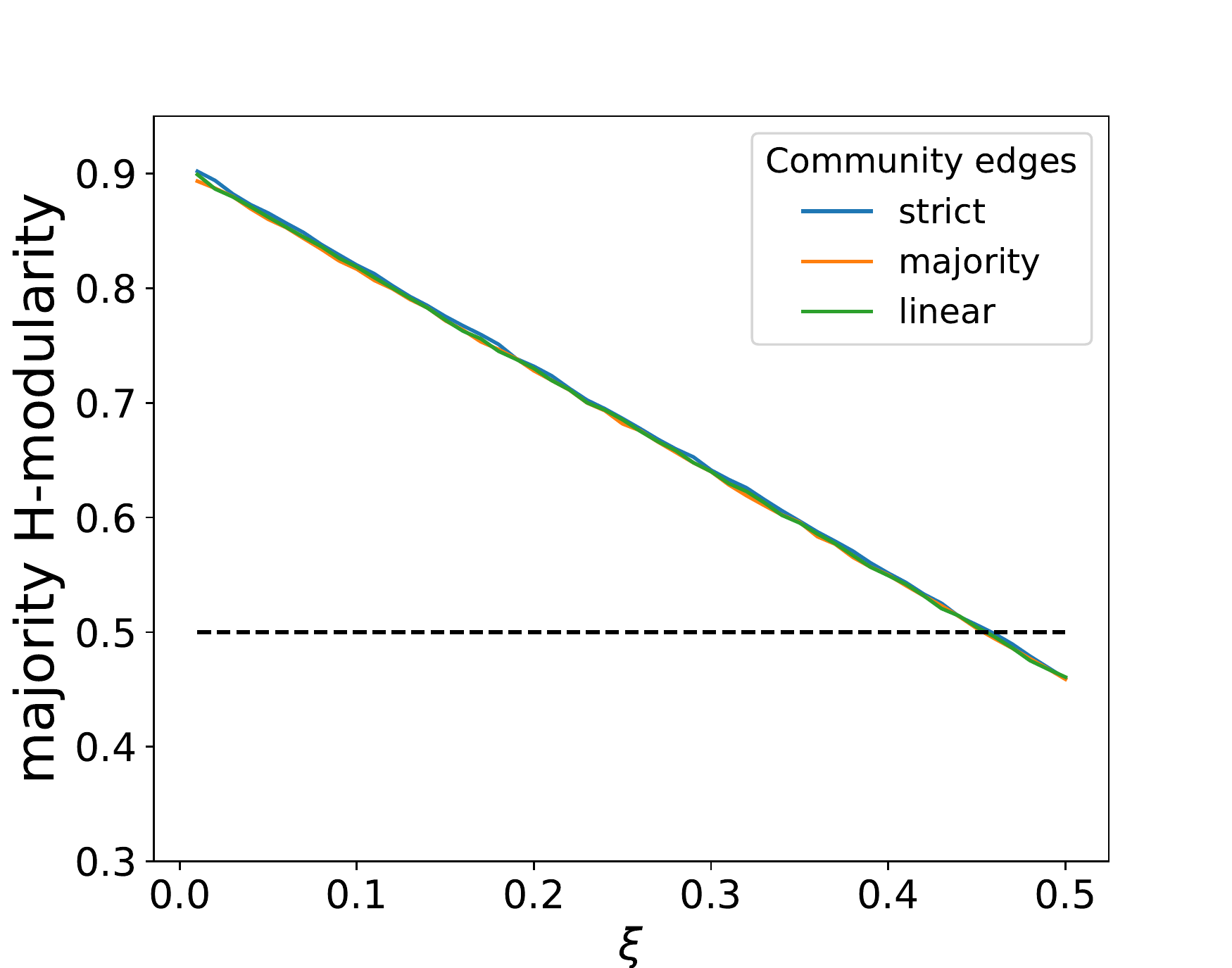}
     \includegraphics[width=0.45\textwidth]{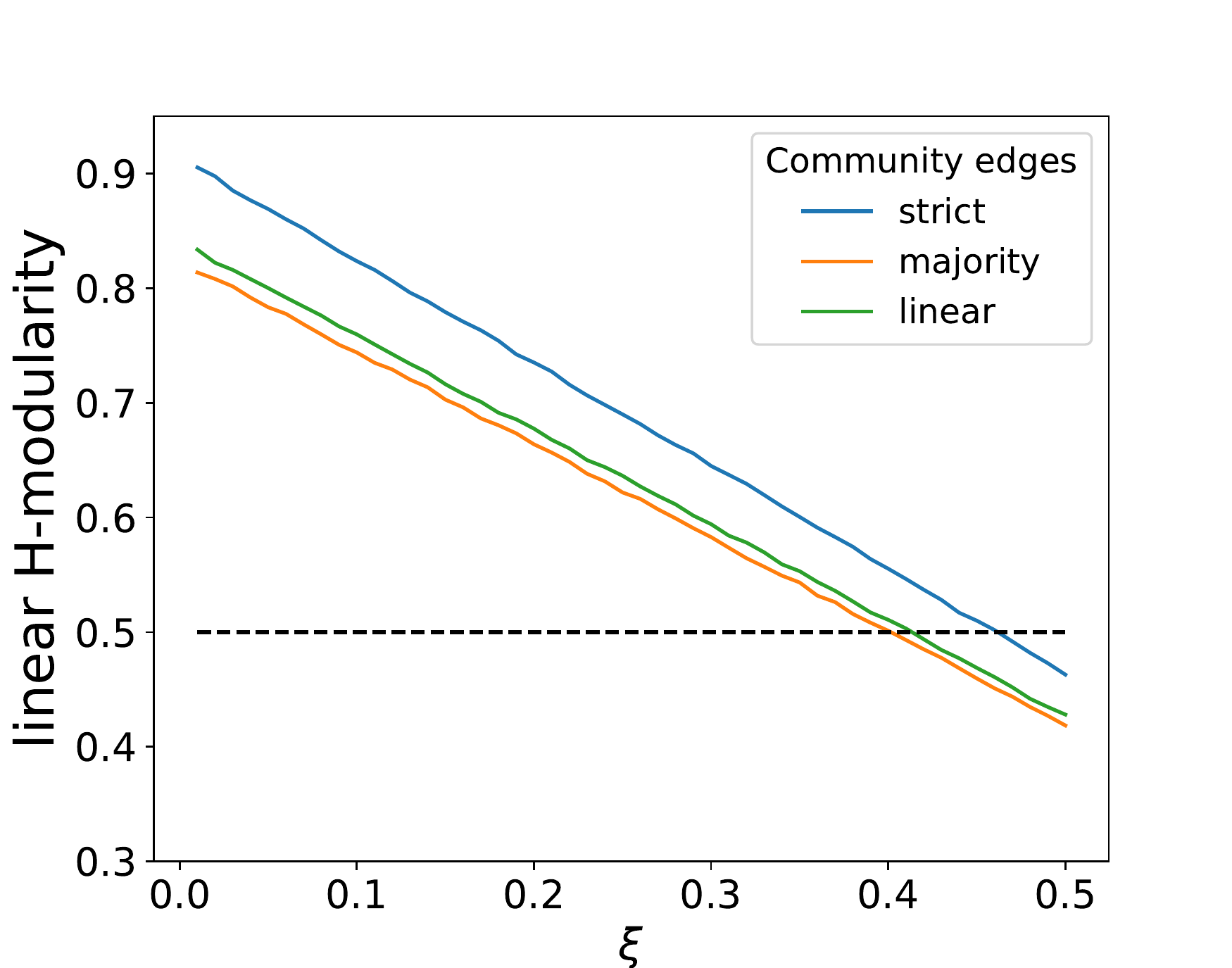}
     \caption{Four modularity functions of \hABCD\ hypergraphs with varying noise parameter $\xi$. The community hyperedges were generated according to three different models: strict, majority, and linear.}
     \label{fig:xis}
\end{figure}

Let us now come back to experiments. We generated hypergraphs on $n=10,000$ nodes and with 50 different choices for the parameter $\xi$ responsible for the level of noise, namely, $\xi \in \{0.01, 0.02, \ldots, 0.50\}$. For each value of $\xi$, three different models for matrix $w$ were tested: majority and strict, that we already experimented with, but also linear---see the beginning of Section~\ref{sec:model} for details. Results are reported in Figure~\ref{fig:xis}. 

Before we start discussing particular figures, let us mention that, in general, for the same level of noise, generating strict hyperedges yields higher modularity, while the majority model yields the smallest; the linear lies in-between the two models. The reason is clear. All models generate the same number of community hyperedges but their levels of homogeneity depends on matrix $w$. Now, the 2-section (graph) modularity replaces each hyperedge with a complete graph and then applies standard graph modularity to the resulted graph. As a result, it favours hyperedges that are more homogeneous since such hyperedges generate more edges within communities in the corresponding 2-section graph and so they contribute more to the modularity. Similarly, each of the three hypergraph modularities value hyperedges that are more homogeneous at least as much as less homogeneous ones, that is, the corresponding parameters $u_{c,d}$ are non-decreasing. In fact, all of them but the majority modularity are strictly increasing. As a consequence, for these modularity functions (see Figure~\ref{fig:xis} top-right and bottom-right), the values for the strict model are larger than for the ones for the linear model which in turn are larger than the values for the majority model. Finally, note that for the majority modularity (see Figure~\ref{fig:xis}, bottom-left), the composition of the community edges does not matter; regardless of which model is used, all community hyperedges contribute equally to the modularity function and so the modularity functions are the same. Another observation is that generating community hyperedges with majority or linear methods yield more similar hypergraphs, while the strict model is visibly distinct.

For each plot in Figure~\ref{fig:xis}, we added a dashed line at the modularity value 0.5. This is to show that with the exception of the majority modularity, different values of the noise parameter $\xi$ are required for different models to obtain the same modularity. It is clear and expected but it is important when the \hABCD\ model is used to benchmark the performance of clustering algorithms so that ``apples and oranges'' are not compared against each other. For example, one can easily generate hypergraphs with similar 2-section modularities using different models for the community hyperedges (strict, majority, linear) and different noise parameters. Such hypergraphs would seem very similar when looking at their two-section graphs, but are really quite different hypergraphs. We show some specific numbers illustrating this phenomena in Table~\ref{tab:twosec}.

\begin{table}
\centering
\begin{tabular}{cc|ccc|c}
\toprule
\multicolumn{2}{c|}{\hABCD} & \multicolumn{4}{c}{Modularity} \\
method &    $\xi$ &      strict &     linear &    majority &   2-section \\
\midrule
strict &  0.43 &  0.533546 &  0.528192 &  0.525261 & 0.501700 \\
linear &  0.25 &  0.514351 & 0.636436 & 0.685292 &    0.504892 \\
majority &  0.20 &  0.508085 &    0.663819 & 0.727940 &  0.502773 \\
\bottomrule
\end{tabular}
\caption{Three hypergraphs with very similar 2-section modularities but different models for the community hyperedges and different noise values. The differences are evident when looking at the corresponding hypergraph modularities.\label{tab:twosec}}
\end{table}

\subsection{Time Complexity of the Algorithm}\label{sec:time_complexity}

The framework to generate ``\ABCD-type'' models is very flexible; the original \ABCD\ model was already parallelized (\textbf{ABCDe}) and generalized to include outliers (\textbf{ABCD+o}). In this paper, we adjust it to hypergraphs (\hABCD). Another important feature of these models is that they are also, by design, very fast. This is in contrast to the main problem with high order structures---there are $\binom{n}{d}$ potential hyperedges of size $d$ which is strikingly larger than the number of potential edges, $\binom{n}{2}$. As a result, hypergraph synthetic graph models are inherently slow but there are some exceptions. For example, the experiments presented in~\cite{ruggeri2022principled} show that their ``model is highly efficient, as it allows to sample sparse hypergraphs of dimensions up to $10^5$ nodes in less than one hour''. Our model generates graphs of order ten times larger in less than 10 seconds. Similarly, \cite{osti_1651312} reports that \textbf{HyGen} takes approximately four minutes to generate a hypergraph with 4.8 million vertices, 1.6 million hyperedges, and 800 clusters using 1,024 processes on a leadership class computing platform. \hABCD\ is able to generate hypergraphs of similar size under one minute on a single core of a desktop computer. This shows a drastic advantage of the used framework, even in comparison to other scalable approaches.

The two most computationally expensive parts of the algorithm are: (1) assignment of nodes to communities, and (2) hyperedge generation (the remaining are sampling node degrees and sampling community sizes and their time only depends on $n$).

The cost of assignment of nodes to communities is $O(n + \lambda\ell L^2)$, where $\lambda \le n$ is the number of unique $(y_i, z_i)$ combinations that appear in the inequality~(\ref{eq:prop_community_assignment}). Note that in power-law graphs there are many nodes of the same degree and so that although $\lambda$ grows with $n$ it is in practice much smaller than it. Similarly, $\ell$ grows with $n$, but is of order of magnitude smaller than it; see~\cite{kaminski2022modularity} for asymptotic analysis of distribution of degree and community sizes.

The cost of hyperedge generation is $O(\vol(V) + \ell L^2)$, where $\ell L^2$ is the cost of preprocessing which has to be done for each community for each $(c,d)$ combination. Again using the results from \cite{kaminski2022modularity} we know that $\vol(V) = O(n)$, and thus we can expect that actual hyperedge generation should be the most expensive part of the algorithm for large values of $n$, provided that $L$ is a fixed constant.

The performance benchmarks results are presented in Table~\ref{tab:timing} to show the scalability of \textbf{h-ABCD} in practice for the case of small values of $L$. In particular, the timings reflect not only asymptotic complexity of the algorithm but also impact on timing of implementation details like: CPU cache locality and memory allocation management cots, as we want to be sure that they do not significantly impact it.

In these experiments, the level of noise was set to $\xi=0.2$, uniform distribution of hyperedge sizes was used (that is, $q_d = 1/(L-1)$ for any $d \in [L] \setminus \{1\}$) together with the majority model. Total time we report, apart from assignment to communities and hyperedge generation includes time to generate node degrees and community sizes (these two times depend only on $n$).

The results of the experiments confirm that hyperedge generation is the most significant component of the hypergraph generation process for the case when $L$ is small compared to $n$. As expected, the number of nodes $n$ has a major impact on total runtime of the algorithm, but also increasing $L$ influences it as is predicted by the asymptotic formulas. In general, the proposed algorithm allows to generate hypergraphs having one million nodes in several seconds in the considered cases.

\medskip

Additionally, in Table~\ref{tab:timing2} we check if the time complexity of the algorithm indeed is proportional to $L^2$. For this test we fix $n=10^6$ and check values of $L$ up to 320, keeping the distribution of hyperedge sizes uniform, that is, $q_d=1/(L-1)$ for $d>1$. For this test we set the larger minimum community size to $s=10,000$. The reason for this change is that communities must be large enough so that hyperedges can fit into the community graphs. The tests show that, indeed, the assignment to communities component becomes the most time consuming as $L$ grows and the relationship is quadratic as predicted by theoretical considerations.

\begin{table}
\centering
\begin{tabular}{rrrrr}
\toprule
$n$ & $L$ & Assignment to communities & Hyperedge generation & Total time* \\
\midrule
500,000 & 5  & 0.12 & 1.67 & 2.09\\
500,000 & 10 & 0.18 & 1.71 & 2.19\\
500,000 & 20 & 0.48 & 1.77 & 2.55\\
\midrule
1,000,000 & 5  &  0.28 & 3.50 & 4.18\\
1,000,000 & 10 &  0.40 & 3.64 & 4.44\\
1,000,000 & 20 & 1.09 & 3.73 & 5.22\\
\midrule
2,000,000 & 5  &  0.70 &  9.77 & 11.48\\
2,000,000 & 10 &  0.87 &  9.88 & 11.74\\
2,000,000 & 20 & 1.97 & 13.81 & 15.76\\
\bottomrule
\multicolumn{5}{p{5.5in}}{* \footnotesize Total time also includes time to generate node degrees and community sizes (these two times depend only on $n$).}
\end{tabular}
\caption{Generation time (in seconds) of \textbf{h-ABCD} hypergraph for different values of $n$ and $L$. \label{tab:timing}}
\end{table}

\begin{table}
\centering
\begin{tabular}{rrrr}
\toprule
$L$ & Assignment to communities & Hyperedge generation & Total time* \\
\midrule
20 & 0.34 & 3.39 & 2.76 \\
40 & 1.07 & 2.18 & 3.28 \\
80 & 8.46 & 3.43 & 11.92 \\
160 & 46.73 & 5.62 & 52.37 \\
320 & 252.78 & 13.19 & 266.01 \\
\bottomrule
\multicolumn{4}{p{5.5in}}{* \footnotesize Total time also includes time to generate node degrees and community sizes.}
\end{tabular}
\caption{Generation time (in seconds) of \textbf{h-ABCD} hypergraph for different values of $L$, with
$n=10^6$ and minimum community size fixed to 10,000. \label{tab:timing2}}
\end{table}

\section{Conclusions}\label{sec:future}

In this paper we introduced \hABCD, one of the very first synthetic random hypergraph models with community structure. This model produces ``\textbf{LFR}-type'' hypergraphs but its building blocks are inherited from the \ABCD\ model. Because of that, one can easily generate synthetic hypergraphs with the degree distribution as well as the distribution of community sizes following power-law. The generation process is fast and the ground truth partition of nodes can be easily used to benchmark hypergraph community detection algorithms. The benchmark is available on GitHub.

Modelling and mining complex networks as hypergraphs is an exciting and brand new research direction. There are many open problems left to be investigated. We plan to work on the following questions next.

\begin{itemize}
\item We plan to design and implement hypergraph clustering algorithm. One approach that is worth considering is to optimize the hypergraph modularity function introduced in~\cite{kaminski2019clustering}; see~\cite{kaminski2020community} for encouraging initial results. In any case, \hABCD\ model will be instrumental in validating and benchmarking various ideas during this process.
\item We plan to analyze typical, asymptotic properties of \hABCD\ model, especially the behaviour of the hypergraph modularity function. Similar study is already done for the original \ABCD\ model in~\cite{kaminski2022modularity} but not all questions are answered.
\item Generating hypergraphs is more challenging and time consuming than generating graphs. \hABCD\ model can generate hypergraphs having 1 million nodes and the maximum hyperedge size 5 on an average laptop in several seconds. It means that the time complexity is similar to the original \ABCD\ graph generator for comparable graph sizes, but significantly faster than the \textbf{LFR} graph generator and much faster than other hypergraph competitors. Since there are still no scalable algorithms for hypergraphs that can handle huge networks, there is no need for parallel and scalable implementations of \hABCD\ yet, but the situation might change in the near future. We plan to implement a multi-threaded version of the model as it was done for the original \textbf{ABCDe} model in~\cite{kaminski2022abcde}.
\item The framework used to generate \ABCD\ model is flexible. It was modified to include outliers (\textbf{ABCD+o}) and in this paper we modified it to generate hypergraphs (\hABCD). Since detecting overlapping communities is an important problem and there are few synthetic benchmarks for this task, it would be good to modify the framework one more time and add option for overlapping communities to our ``tool-box''.
\item \hABCD\ model allows for any value of $L$, the maximum size of hyperedges, but it is designed with relatively small values of $L$ in mind such as $L=10$ or $L=20$. Such hypergraphs are typically of interest in the context of community detection algorithms. However, if one wants to create hypergraphs with much larger values of $L$, then the algorithm will slow down. Indeed, ``collisions'' when generating large hyperedges will occur with larger probability. In order to handle such large hyperedges, the algorithm needs to be slightly adjusted. The drawback, and the reason why we do not do it currently, is that the distribution of hyperedges would deviate from uniform distribution.
\end{itemize}

\section*{Acknowledgements}
PP and BK have been supported by the Polish National Agency for Academic Exchange under the Strategic Partnerships programme, grant number BPI/PST/2021/1/00069/U/00001.

\bibliography{ref}


\end{document}